\documentclass[sigconf]{acmart}

\AtBeginDocument{%
  \providecommand\BibTeX{{%
    \normalfont B\kern-0.5em{\scshape i\kern-0.25em b}\kern-0.8em\TeX}}}

\usepackage{comment}

\usepackage{hyperref}
\usepackage{url}
\usepackage{xurl}
\usepackage[utf8]{inputenc} 
\usepackage[T1]{fontenc}    
\usepackage{hyperref}       
\usepackage{booktabs}       
\usepackage{amsfonts}       
\usepackage{nicefrac}       
\usepackage{microtype}      
\usepackage{xcolor}
\usepackage{microtype}
\usepackage{graphicx}
\usepackage{booktabs} 
\usepackage{paralist}
\usepackage{amsmath}
\usepackage{amsfonts}
\usepackage{caption}
\usepackage{subcaption}
\usepackage{xcolor}
\usepackage{colortbl}
\usepackage{hyperref}
\usepackage{amsmath}
\usepackage{mathtools}
\usepackage{amsthm}
\usepackage[capitalize,noabbrev]{cleveref}
\usepackage{wrapfig}
\usepackage{makecell}
\usepackage{multirow}
\usepackage{booktabs,hhline}
\usepackage{pifont}
\usepackage{algorithm}
\usepackage{algorithmic}
\usepackage{multicol}
\usepackage{natbib}
\usepackage{tabularx}
\usepackage{clipboard}
\usepackage{float}
\usepackage{tablefootnote}

\newcommand{\ie}{\textit{i}.\textit{e}.}
\newcommand{\eg}{\textit{e}.\textit{g}.}
\newcommand{\etc}{\textit{etc}. }
\definecolor{mygreen}{RGB}{0, 0, 0}
\definecolor{cellgreen}{HTML}{C6EFCE}
\definecolor{cellyellow}{HTML}{FFEB9C}
\definecolor{cellred}{HTML}{FFC7CE}

\newcommand{\todo}[1]{\textcolor{red}{\textbf{\small{[TODO: #1]}}}}

\theoremstyle{plain}

\theoremstyle{definition}

\theoremstyle{remark}

\copyrightyear{2023}
\acmYear{2023}
\setcopyright{acmlicensed}\acmConference[WWW '23]{Proceedings of the ACM Web Conference 2023}{April 30-May 4, 2023}{Austin, TX, USA}
\acmBooktitle{Proceedings of the ACM Web Conference 2023 (WWW '23), April 30-May 4, 2023, Austin, TX, USA}
\acmPrice{15.00}
\acmDOI{10.1145/3543507.3587430}
\acmISBN{978-1-4503-9416-1/23/04}

\begin{document}

\title{A Prompt Log Analysis of Text-to-Image Generation Systems}



\author{Yutong Xie}
\authornote{These authors contributed equally to this research.}
\email{yutxie@umich.edu}
\affiliation{%
  \institution{University of Michigan}
  \city{Ann Arbor}
  \state{Michigan}
  \country{USA}
}


\author{Zhaoying Pan}
\authornotemark[1]
\email{panzy@umich.edu}
\affiliation{%
  \institution{University of Michigan}
  \city{Ann Arbor}
  \state{Michigan}
  \country{USA}
}

\author{Jinge Ma}
\authornotemark[1]
\email{jingema@umich.edu}
\affiliation{%
  \institution{University of Michigan}
  \city{Ann Arbor}
  \state{Michigan}
  \country{USA}
}

\author{Luo Jie}
\email{rogerluo@nianticlabs.com}
\affiliation{%
  \institution{Niantic, Inc.}
  \city{San Francisco}
  \state{California}
  \country{USA}
}

\author{Qiaozhu Mei}
\email{qmei@umich.com}
\affiliation{%
  \institution{University of Michigan}
  \city{Ann Arbor}
  \state{Michigan}
  \country{USA}
}


\begin{abstract}
Recent developments in large language models (LLM) and generative AI have unleashed the astonishing capabilities of text-to-image generation systems
to synthesize high-quality images that are faithful to a given reference text, known as a ``prompt''. These systems have immediately received lots of attention from researchers, creators, and common users. Despite the plenty of efforts to improve the generative models, there is limited work on understanding the information needs of the users of these systems at scale. We conduct the first comprehensive analysis of large-scale prompt logs collected from multiple text-to-image generation systems. Our work is analogous to analyzing the \emph{query logs} of Web search engines, a line of work that has made critical contributions to the glory of the Web search industry and research. 
Compared with Web search queries, text-to-image prompts are significantly longer, often organized into special structures that consist of the \emph{subject}, \emph{form}, and \emph{intent} of the generation tasks and present unique categories of information needs. 
Users make more edits within creation sessions, which present remarkable \emph{exploratory} patterns. 
There is also a considerable gap between the user-input prompts and the captions of the images included in the open training data of the generative models. 
Our findings provide concrete implications on how to improve text-to-image generation systems for creation purposes.
\end{abstract}





\received{7 February 2023}
\received[revised]{15 March 2023}
\received[accepted]{6 March 2023}

\maketitle

\section{Introduction}
\label{sec:introduction}

Recent developments in large language models (LLM) (e.g., GPT-3 \cite{brown2020language}, PaLM \cite{chowdhery2022palm}, LLaMA \cite{touvron2023llama}, and GPT-4 \cite{gpt4}) and generative AI (especially the diffusion models \cite{sohl2015deep,ho2020denoising}) have enabled the astonishing image synthesis capabilities of text-to-image generation systems, such as DALL·E \cite{ramesh2021zero,ramesh2022hierarchical}, Midjourney \cite{midjourney}, latent diffusion models (LDMs) \cite{rombach2022high}, Imagen \cite{saharia2022photorealistic}, and Stable Diffusion \cite{rombach2022high}.
As these systems are able to produce images of high quality that are faithful to a given reference text (known as a ``prompt''), they have immediately become a new source of creativity \cite{oppenlaender2022creativity} and attracted a great number of creators, researchers, and common users. As a major prototype of generative AI, many believe that these systems are introducing fundamental changes to the creative work of humans \cite{davenport_mittal_2022}.

Despite plenty of efforts on improving the performance of the underneath generative models, there is limited work on analyzing the information needs of the real users of these text-to-image systems, regardless of the cruciality to understand the objectives and workflows of the creators and identify the gaps in how the current systems are capable of facilitating the creators' needs. 

In this paper, we take the initiative to investigate the information needs of text-to-image generation by conducting a comprehensive analysis of millions of user-input prompts in multiple popular systems, including Midjourney, Stable Diffusion, and LDMs. Our analysis is analogous to \emph{query log analysis} of search engines, a line of work that has inspired many developments of modern information retrieval (IR) research and industry   \cite{jansen1998real,silverstein1999analysis,broder2002taxonomy,herskovic2007day,yang2011query}. 
In this analogy, a text-to-image generation system is compared to a search engine, the pretrained large language model is compared to the search index, a user-input prompt can be compared to a search query that describes the user's information need, while a text-to-image generation model can be compared to the search or ranking algorithm that generates (rather than retrieves) one or multiple pieces of content (images) to fulfill the user's need (Table \ref{tab:analogy}). 

\begin{table}[htbp]
    \vspace{15pt}
    \caption{The analogy between text-to-image generation and Web or vertical search. }
    \vspace{-5pt}
    \label{tab:analogy}
    \centering
    \begin{tabular}{c|c}
        \toprule
        \textbf{Text-to-image generation} & \textbf{Web and vertical search} \\
        \midrule
        Images & Webpages/Documents \\
        Generation & Retrieval \\
        Text-to-image generation system & Search engine \\
        Prompt & Query \\
        Pretrained language model & Search index \\
        Image generation models & Ranking algorithms \\
        Prompt log analysis & Query log analysis \\
        $\dots$ & $\dots$ \\
        \bottomrule
    \end{tabular}
    \vspace{-10pt}
\end{table}

Through a large-scale analysis of the prompt logs, we aim to answer the following questions:
\Copy{four-questions}{
\textcolor{mygreen}{
\begin{inparaenum}[(1)]
    \item \emph{How do users describe their information needs in the prompts}?
    \item \emph{How do the information needs in text-to-image generation compare with those in Web search}?
    \item \emph{How are users' information needs satisfied}?
    \item \emph{How are users' information needs covered by the image captions in open datasets}?
\end{inparaenum}
}
}

The results of our analysis suggest that 
\begin{inparaenum}[(1)]
    \item text-to-image prompts are usually structured with terms that describe the \emph{subject}, the \emph{form}, and the \emph{intent} of the image to be created (Sec. \ref{sec:prompts-structure});
    \item text-to-image prompts are sufficiently different from Web search queries. Besides significantly lengthier prompts and sessions, there is especially a prevalence of \emph{exploratory} prompts (Sec. \ref{sec:prompts-queries});
    \item image generation quality (measured by user rating) is correlated with the length of the prompt as well as the usage of terms (Sec. \ref{sec:prompts-ratings}); and 
    \item there is a considerable gap between the user-input prompts and the image captions in open datasets (Sec. \ref{sec:prompt-train-diff}). 
\end{inparaenum}
\textcolor{mygreen}{
More details of our analysis are listed in the Appendix, and the code and the complete results are accessible via our GitHub repository\footnote{ GitHub repository: \url{https://github.com/zhaoyingpan/prompt_log_analysis}.}. 
}
Based on these findings, we conclude several challenges and actionable opportunities of text-to-image generation systems (Sec. \ref{sec:challenges}). 
We anticipate our study would help the text-to-image generation community to better understand and facilitate creativity on the Web. 



\section{Related Work}

\subsection{Text-to-Image Generation}

Text-to-Image generation is a multi-modal task that aims to translate text descriptions (known as ``prompts'') into faithful images of high quality. 
Recent text-to-Image generation models could be categorized into two main streams:  
\begin{inparaenum}[(1)]
    \item models based on variational autoencoders (VAEs) \cite{kingma2013auto} or generative adversarial networks (GANs) \cite{NIPS2014_5ca3e9b1}, and 
    \item models built upon denoising diffusion probabilistic models (DDPMs, or diffusion models) \cite{ho2020denoising}.
\end{inparaenum}

The earliest application of deep neural networks in text-to-image generation could be dated back to 2015, where \citet{mansimov2015generating} 
proposed to generate images from texts using a recurrent VAE with the attention mechanism. 
In the next few years, \citet{reed2016generative} and \citet{Cho2020XLXMERT} started to use GANs as generative models from texts to images. 
These models have made it possible to generate images from texts, however, most generated images are blurry and consist of simple structures. 
Later in 2021, OpenAI released DALL·E, combining the powerful GPT-3 language model \cite{brown2020language} as the text encoder and a VAE as the image generator \cite{ramesh2021zero}. 
DALL·E is able to generate more complex and realistic images, establishing a new standard of text-to-image generation.

Since late 2021, with the advances of DDPMs (diffusion models), several compelling text-to-image generation systems have been developed and released to the public,
demonstrating astounding capabilities in faithful image synthesis and bringing text-to-image generation into a new era. 
They include Disco Diffusion\footnote{Disco Diffusion: \url{https://github.com/alembics/disco-diffusion}, retrieved on 3/14/2023.}, GLIDE \cite{nichol2021glide}, Midjourney \cite{midjourney}, DALL$\cdot$E 2 \cite{ramesh2022hierarchical}, latent diffusion models (LDMs) \cite{rombach2022high}, Imagen \cite{saharia2022photorealistic} and Stable Diffusion \cite{rombach2022high}.
These systems immediately become a trend in the art creation community, attracting both artists and common users to create with such systems \cite{oppenlaender2022creativity}. 


\vspace{-5pt}
\subsection{Text-to-Image Prompt Analysis}

Despite plenty of efforts on improving the performance of the underneath generative models, there is limited work on analyzing the user-input prompts and understanding the information needs of the real users of text-to-image systems. 

\citet{liu2022design} explored what prompt keywords and model hyperparameters can lead to better generation performance from the human-computer interaction (HCI) perspective. In particular, 51 keywords related to the subject and 51 related to the style have been tested through synthetic experiments. 
\citet{oppenlaender2022taxonomy} further conducted an autoethnographic study on the modifiers in the prompts. As a result, six types of prompt modifiers have been identified, including subject terms, style modifiers, image prompts, quality boosters, repetition, and magic terms. 
In addition, \citet{pavlichenko2022best} presented a human-in-the-loop approach and extracted some most effective combinations of prompt keywords. 

These studies have provided valuable insights into certain aspects of text-to-image prompts. However, these researches are mostly based on small numbers of independent prompts and/or computational experiments. These lab experiments usually do not consider prompts in real usage sessions and can hardly reflect the ``whole picture'' of the information needs of the real users. 
Our study provides the first large-scale quantitative analysis based on the user inputs collected from real systems. 
Besides, we also compare the characteristics of prompts with those of Web search queries as well as image captains in open text-image datasets, revealing considerable differences and practical implications. 

\subsection{Query Log Analysis}

Query log analysis of Web search engines is a classical line of work that has inspired many developments in modern information retrieval (IR) research and industry. 
Such an analysis usually includes examinations into terms, queries, sessions, and users \cite{jansen1998real,silverstein1999analysis,broder2002taxonomy}. 
Aside from the general research on Web search engines, query log analysis has also been conducted on vertical search engines like medical search engines (\eg, PubMed and electronic health records (EHR) search engine), where the analysis results are further compared with Web search patterns \cite{herskovic2007day,yang2011query}. 

In this paper, we make an analogy between query log analysis and prompt analysis. In this analogy, a user-input prompt can be compared to a \emph{search query} that describes the user's information need, while a text-to-image generation system could be compared to a \emph{search engine} that generates (rather than retrieves) one or more pieces of content (in our case, the image(s)) to fulfill the user's need. 

\vspace{-2pt}
\section{Prompt Log Datasets}
\label{sec:data}

We consider three large and open prompt log datasets, including the Midjourney Discord dataset \cite{iuliaturc_gauravnemade_2022}, the DiffusionDB \cite{wangDiffusionDBLargescalePrompt2022}, and the Simulacra Aesthetic Captions (SAC) \cite{pressmancrowson2022}. These datasets involve three popular text-to-image generation systems -- Midjourney \cite{midjourney}, Stable Diffusion \cite{rombach2022high}, and latent diffusion models (LDMs) \cite{rombach2022high}. 

\paragraph{\textbf{The Midjourney Discord dataset}} The Midjourney dataset \cite{iuliaturc_gauravnemade_2022} is obtained by crawling message records from the Midjourney Discord community over a period of four weeks (June 20 -- July 17, 2022). This dataset contains approximately 250K records, with user-input prompts, URLs of generated images, usernames, user IDs, message timestamps, and other Discord message metadata.

\paragraph{\textbf{DiffusionDB}} DiffusionDB \cite{wangDiffusionDBLargescalePrompt2022} is a large-scale dataset with 14M images generated with Stable Diffusion \cite{rombach2022high}. For each image, this dataset also provides the corresponding prompt, user ID, timestamp, and other meta information. 


\paragraph{\textbf{Simulacra Aesthetic Captions (SAC)}} The SAC dataset \cite{pressmancrowson2022} contains 238K images generated from over 40K user-submitted prompts with LDMs \cite{rombach2022high}.  
SAC annotates images with aesthetic ratings in the range of $[1,10]$ collected from surveys. The prompts in SAC are also relatively clean. However, SAC does not include information about user IDs or timestamps. 

Table \ref{tab:dataset-stats} lists the basic statistics of the datasets. In the raw data, one input prompt can correspond to multiple generated images and create multiple data entries for the same input. We remove these duplicates while reserving repeated inputs from users.
More details about the data and data processing are described in Appendix \ref{app:data}. 

\begin{table}[t]
\caption{Statistics of datasets. The values (except for the raw number of records) are calculated after data processing. }
\vspace{-5pt}
\label{tab:dataset-stats}
\resizebox{\columnwidth}{!}{
\begin{tabular}{l|rrr}
\toprule
\textbf{Dataset}	&	\textbf{Midjourney}	& \textbf{DiffusionDB} 	& \textbf{SAC} \\	
\midrule					
\textbf{Raw \#Records}	&	250K	& 14M	& 238K\\
\textbf{\#Prompts}	&	145,074	& 2,208,019	& 34,190 	\\
\textbf{\#Unique prompts} & 122,905 & 1,817,721 & 34,190 \\
\textbf{\#Unique terms}	&	97,052	&  182,386	&  22,898 	\\
\midrule
\textbf{\#Users}	&	1,665	&  10,380	& N/A \\
\textbf{Median \#prompts/user} & 12     & 62     & N/A \\
\textbf{Max \#prompts/user}    & 2,493  & 19,556 & N/A 	\\
\bottomrule
\end{tabular}}
\vspace{-10pt}
\end{table}

\section{Prompt Log Analysis}
\label{sec:analysis}

We analyze the prompts in the datasets and aim to answer the four questions mentioned in Section~\ref{sec:introduction}.

\textcolor{mygreen}{
\subsection{How do Users Describe Information Needs?}
}
\label{sec:prompts-structure}

We first 
\textcolor{mygreen}{
investigate how users describe their information needs by exploring
}
the structures of prompts. 
We start with analyzing the usage of terms (tokens or words) in prompts. 
We conduct a first-order analysis that focuses on term frequency, 
followed by a second-order analysis that focuses on co-occurring term pairs. The significance of a term pair is measured with the $\chi^2$ metric \cite{agresti2012categorical,silverstein1999analysis}: 

\vspace{-7pt}
\begin{small}
\begin{eqnarray}
\chi^2(a, b) & = & \ \frac{[E(a b)-O(a b)]^2}{E(a b)}+\frac{[E(\bar{a} b)-O(\bar{a} b)]^2}{E(\bar{a} b)}+ 
\nonumber
\\ 
& & \ \frac{[E(a \bar{b})-O(a \bar{b})]^2}{E(a \bar{b})}+\frac{[E(\bar{a} \bar{b})-O(\bar{a} \bar{b})]^2}{E(\bar{a} \bar{b})},
\end{eqnarray}
\end{small}
where $a,b$ are two terms, $O(ab)$ is the number of prompts they co-occur in, $E(ab)$ is the expected co-occurrences under the independence assumption, and $\bar{a}$, $\bar{b}$ stand for the absence of $a$, $b$.




\begin{table}[htbp]
\caption{Most frequent terms used in prompts. }
\vspace{-5pt}
\label{tab:terms-first}
\resizebox{\columnwidth}{!}{
\begin{tabular}{r|lr|lr|lr}
\toprule
   & \multicolumn{2}{c|}{\textbf{Midjourney}}                                & \multicolumn{2}{c|}{\textbf{DiffusionDB}}                                       & \multicolumn{2}{c}{\textbf{SAC}}                               \\
   \hhline{~------}
   & \textbf{Term}      & \textbf{Freq.}     & \textbf{Term}  & \textbf{Freq.}  & \textbf{Term}      & \textbf{Freq.}      \\
\midrule
1  & ,                   & 91,993     & ,                & 1,689,552       & ,              & 17,265  \\
2  & \textgreater{}      & 71,353     & of               & 1,084,836       & of             & 15,400  \\
3  & \textless{}         & 71,348     & a                & 1,043,542       & a              & 14,693  \\
4  & of                  & 56,470     & by               & 943,503        & by             & 12,442  \\
5  & a                   & 47,987     & and              & 720,802        & the            & 9,319   \\
6  & in                  & 42,685     & in               & 669,648        & and            & 7,614   \\
7  & -{}-ar                & 40,014     & detailed         & 653,587        & in             & 7,186   \\
8  & the                 & 38,155     & art              & 598,493        & on             & 6,723   \\
9  & and                 & 33,330     & the              & 572,569        & artstation     & 5,935   \\
10 & by                  & 28,074     & artstation       & 484,898        & .              & 5,686   \\
11 & detailed            & 25,134     & on               & 475,406        & portrait       & 5,652   \\
12 & with                & 24,112     & painting         & 426,930        & art            & 5,598   \\
13 & style               & 23,461     & portrait         & 412,547        & with           & 4,444   \\
14 & on                  & 20,282     & with             & 402,008        & painting       & 4,347   \\
15 & render              & 20,061     & highly           & 334,410        & illustration   & 3,359   \\
16 & cinematic           & 19,782     & k                & 320,290        & -              & 3,354   \\
17 & 16:9                & 18,616     & lighting         & 310,598        & oil            & 3,234   \\
18 & realistic           & 18,012     & digital          & 310,336        & concept        & 3,214   \\
19 & -                   & 17,677     & -                & 287,732        & digital        & 2,997   \\
20 & octane              & 16,925     & intricate        & 276,246        & beautiful      & 2,695   \\ 

\bottomrule
\end{tabular}}
\vspace{-10pt}
\end{table}



In Table \ref{tab:terms-first}, we list the most frequent terms, measured by the number of text-to-image prompts they appear in. The most significant term pairs are listed in Table \ref{tab:terms-second}. Based on the first- and second-order analysis results, we present the following findings. 




\begin{table}[htbp]
\caption{Most significant term pairs used in the same prompt.}
\vspace{-5pt}
\label{tab:terms-second}
\resizebox{\columnwidth}{!}{
\begin{tabular}{r|lr|lr|lr}
\toprule
   & \multicolumn{2}{c|}{\textbf{Midjourney}} & \multicolumn{2}{c|}{\textbf{DiffusionDB}}     & \multicolumn{2}{c}{\textbf{SAC}}     \\
\hhline{~------}
   & \textbf{Pair}                      & $\chi^2$  & \textbf{Pair}           & $\chi^2$ & \textbf{Pair}          & $\chi^2$ \\
\midrule
1	&	(norman, rockwell)	&	0.285	&	(donald, trump)	&	0.345	&	(matsunuma, shingo)	&	1.000	\\
2	&	(fenghua, zhong)	&	0.250	&	(emma, watson)	&	0.321	&	(lisa, mona)	&	1.000	\\
3	&	(ngai, victo)	&	0.240	&	(biden, joe)	&	0.283	&	(elon, musk)	&	1.000	\\
4	&	(makoto, shinkai)	&	0.237	&	(shinkawa, yoji)	&	0.266	&	(ariel, perez)	&	1.000	\\
5	&	(ray, trace)	&	0.125	&	(blade, runner)	&	0.255	&	(angeles, los)	&	1.000	\\
6	&	(fiction, science)	&	0.123	&	(katsuhiro, otomo)	&	0.238	&	(bradley, noah)	&	1.000	\\
7	&	(anderson, wes)	&	0.106	&	(contest, winner)	&	0.237	&	(hayao, miyazaki)	&	1.000	\\
8	&	(11:17, circa)	&	0.074	&	(takato, yamamoto)	&	0.236	&	(finnian, macmanus)	&	1.000	\\
9	&	(jia, ruan)	&	0.071	&	(``, '')	&	0.216	&	(bartlett, bo)	&	1.000	\\
10	&	(cushart, krenz)	&	0.070	&	(mead, syd)	&	0.130	&	(hasui, kawase)	&	0.500	\\
11	&	(shinkawa, yoji)	&	0.062	&	(akihiko, yoshida)	&	0.123	&	(daniela, uhlig)	&	0.332	\\
12	&	(albert, bierstadt)	&	0.060	&	(elvgren, gil)	&	0.114	&	(edlin, tyler)	&	0.318	\\
13	&	(katsuhiro, otomo)	&	0.057	&	(new, york)	&	0.114	&	(jurgens, mandy)	&	0.286	\\
14	&	({[}, {]})	&	0.053	&	(gi, jung)	&	0.106	&	(bacon, francis)	&	0.286	\\
15	&	(annie, leibovitz)	&	0.052	&	(dore, gustave)	&	0.103	&	(araki, hirohiko)	&	0.258	\\
16	&	(adams, ansel)	&	0.045	&	(star, wars)	&	0.092	&	(radke, scott)	&	0.257	\\
17	&	(mignola, mike)	&	0.043	&	(fiction, science)	&	0.087	&	(ca', n't)	&	0.252	\\
18	&	(1800s, tintype)	&	0.036	&	(league, legends)	&	0.082	&	(card, tarot)	&	0.201	\\
19	&	(dore, gustave)	&	0.036	&	(rule, thirds)	&	0.074	&	(claude, monet)	&	0.190	\\
20	&	(adams, tintype)	&	0.029	&	(ngai, victo)	&	0.061	&	(gogh, van)	&	0.180	\\
\bottomrule
\end{tabular}}
\vspace{-10pt}
\end{table}

\subsubsection{\textbf{Words in prompts describe subjects, forms, and intents. }} 
\label{sec:art-components}

In Art, a piece of work is typically described with three basic components: \emph{subject}, \emph{form}, and \emph{content}. 
In general, the \emph{subject} defines ``what'' (the topic or focus); the \emph{form} confines ``how''  (the development, composition, or substantiation); and the \emph{content} articulates ``why'' (the intention or meaning) \citep{ocvirk1968art}. We are able to relate terms in a text-to-image prompt to these three basic components. Note that the \emph{subject}, \emph{form}, and \emph{content} of a work of art is often intertwined with each other. For example, a term describing the \emph{subject} might also be related to the \emph{form} or \emph{content} and vice versa.  

\paragraph{Subject}

A prompt often contains terms describing its topic or focus, referred to as the \emph{subject}, which can be a person, an object, or a theme \cite{oppenlaender2022taxonomy,pavlichenko2022best}. Among the 50 most frequent terms of all three datasets (parts of them listed in Table \ref{tab:terms-first}), we discover 9 terms related to the \emph{subject}: 
``portrait’’, ``lighting’’, ``light’’, ``face’’, ``background’’, ``character’’, ``man’’, ``head’’, and ``space’’. More examples can be found in Table \ref{tab:terms-second}, such as (``donald'', ``trump''), (``emma'', ``watson''), (``biden'', ``joe''), (``elon'', ``musk''), (``mona'', ``lisa''), (``new'', ``york''), (``los'', ``angeles''), (``star'', ``wars''), (``league'', ``legends''), and (``tarot'', ``card'').

\paragraph{Form}

The \emph{form} confines the way in which an artwork is organized, referring to the use of the \emph{principles of organization} to arrange the elements of art. 
These elements may include \textit{line}, \textit{texture}, \textit{color}, \textit{shape}, and \textit{value}; while the principles of organization consider \textit{harmony}, \textit{variety}, \textit{balance}, \textit{proportion}, \textit{dominance}, \textit{movement}, and \textit{economy}, \etc \cite{ocvirk1968art}. 
Comparably, the \emph{form} of a prompt is usually described as \textit{constraints} to image generation \cite{oppenlaender2022taxonomy,pavlichenko2022best}. Among the top 50 terms of all datasets (parts of them listed in Table \ref{tab:terms-first}), we find 25 terms that are \emph{form}-related: 
``detailed''/``detail’’’, ``art’’, ``painting’’, ``style’’, ``render’’, ``illustration’’, ``cinematic’’, ``k’’ (\eg, ``4K’’ or ``8K’’), ``16:9’’/``9:16’’, ``oil’’ (\eg, ``oil painting’’), ``realistic’’, ``concept’’ (\eg, ``concept art’’), ``digital’’, ``intricate’’, ``black’’, ``dark’’, ``unreal’’, ``white’’, ``sharp’’, ``fantasy’’, ``photo’’, ``smooth’’, and ``canvas’’. 

In addition to these terms, we also notice names of art community Websites (\eg, ArtStation\footnote{ArtStation: \url{https://www.artstation.com/}, retrieved on 3/14/2023. }, Artgerm\footnote{Artgerm: \url{https://artgerm.com/}, retrieved on 3/14/2023. }, and CGSociety\footnote{CGSociety: \url{https://cgsociety.org/}, retrieved on 3/14/2023.}), rendering engines (\eg, Unreal Engine\footnote{Unreal Engine: \url{https://www.unrealengine.com/}, retrieved on 3/14/2023. } and OctaneRender\footnote{OctaneRender: \url{https://home.otoy.com/render/octane-render/}, retrieved on 3/14/2023. }), and artists (\eg, wlop, Norman Rockwell, Fenghua Zhong, Victo Ngai, Shingo Matsunuma, Claude Monet, and Van Gogh, \etc) that appear frequently in the prompts (Tables \ref{tab:terms-first}-\ref{tab:terms-second}). These terms are often used to constrain the style of images, so can be interpreted as \textit{form}-related. 

\paragraph{Intent}

The \emph{content} (as defined in the Art literature) of a prompt tells the intention or purpose of the user and is often described as the emotional or intellectual message that the user wants to express. Among the three components of art, the \emph{content} is the most abstract and can be difficult to identify \cite{ocvirk1968art}. To avoid ambiguity (``content'' has specific meanings in the Web and the AI literature), we name this component of a prompt the ``\textit{intent}'' instead. In the top 50 terms of all datasets (parts of them listed in Table \ref{tab:terms-first}), we find only three terms that might be related to the \emph{intent}: ``beautiful’’, ``trending’’, and ``featured’’. If we go down the list, we are able to identify more: ``epic'', ``moody'', ``fantasy'', ``dramatic'', ``masterpiece'', \etc 

\paragraph{Other terms} 

Aside from the terms that describe the \emph{subject}, \emph{form}, and \emph{intent}, other types of frequently used terms include punctuations (\eg, ``,'' and ``.''), model-specific syntactic characters (\eg, ``<'', ``>'', ``-{}-ar'', and ``::'' that specify model parameters in the Midjourney dataset), and stop words (\eg, ``of'', ``the'', ``in'', ``a'', ``and'', and ``by''). 

\bigbreak

Overall, we find that many of the prompts consist of one or more blocks of terms in at least one of these three categories. The frequent appearance of \emph{form}-related terms is particularly interesting, which adds constraints to the creation process. Future developments of text-to-image generation should consider how to optimize for the users' intents under these constraints. 

\subsubsection{\textbf{Prompts indicate potential applications. }}\label{sec:applications} 

In the second-order analysis, we also discover interesting combinations of terms that might indicate potential applications of text-to-image generation in various areas:

\begin{itemize}
    \item \emph{Image processing}: (``film'', ``grain''), (``blur'', ``blurry''), (``iso'', ``nikon''), (``hdr'', ``professional''), (``flambient'', ``professional''), (``post'', ``processing''), (``color'', ``scheme''), \etc
    \item \emph{Image rendering}: (``ray'', ``trace/tracing''), (``fluid'', ``redshift''), (``unreal'', ``engine''), (``3d'', ``shading''), (``3d'', rendering''), (``global'', ``illumination''), (``octane'', ``render''), \etc
    \item \emph{Graphic design}: (``movie'', ``poster''), (``graphic'', ``design''), (``key'', ``visual''), (``cel'', ``shaded''), (``comic'', ``book''), (``anime'', ``visual''), (``ghibli'', ``studio''), (``disney'', ``pixar''), \etc
    \item \emph{Industrial design}: (``circuit'', ``boards''), (``sports'', ``car''), \etc 
    \item \emph{Fashion design}: (``fashion'', ``model''), (``curly'', ``hair''), \etc
\end{itemize}
These pairs are often related to the \emph{forms} or/and \emph{intents} of the creation, indicating considerable opportunities to develop customized applications for different forms and intents of creative activities.

\subsection{How do Text-to-Image Prompts Compare with Web Search
Queries?}
\label{sec:prompts-queries}

A text-to-image prompt is analogous to a \emph{query} submitted to a Web search engine (image generation model) that retrieves (generates) documents (images) that satisfy the information need (Table \ref{tab:analogy}).  
It is intriguing to compare text-to-image prompts with Web search queries to further understand their similarities and differences. 

\subsubsection{\textbf{Term frequencies do not follow the power law.} } 

\begin{figure}[t]
    \centering
    \includegraphics[width=1\linewidth]{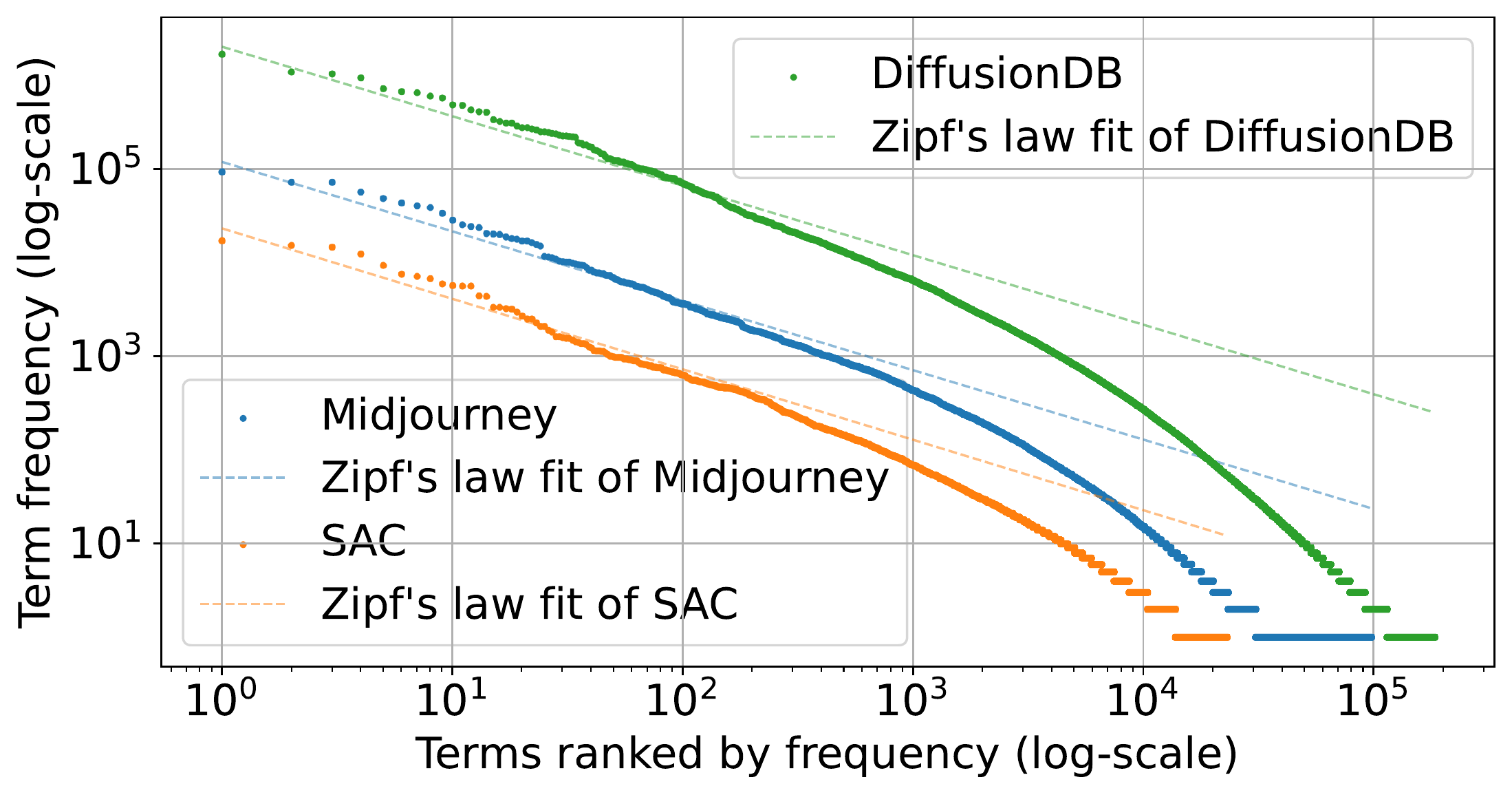}
    \vspace{-15pt}
    \caption{Term frequencies in a log-log scale. The distributions deviate from Zipf's law with exponential tails. 
    }
\label{fig:term-freq}
\end{figure}

While a power law distribution (or a Zipf's distribution when the rank of terms is the independent variable) of term frequency is commonly observed in large-scale corpora and Web search queries \cite{xie2002locality},
we find that the distribution of terms in text-to-image prompts deviates from this pattern. 
Figure~\ref{fig:term-freq} shows that the frequencies of top-ranked terms present a milder decay than Zipf's law, and the tail terms present a clear exponential tail \cite{clauset2009power}. This is likely due to the specialized nature of creative activities, where the use of terms is more restricted than open Web search. This indicates the opportunity and feasibility of curating specialized vocabularies for creation, something similar to the Unified Medical Language System (UMLS) in the biomedical and health domain \cite{bodenreider2004unified}.

\subsubsection{\textbf{Prompt frequencies follow the power law. }}

\begin{figure}[t]
    \centering
    \includegraphics[width=\linewidth]{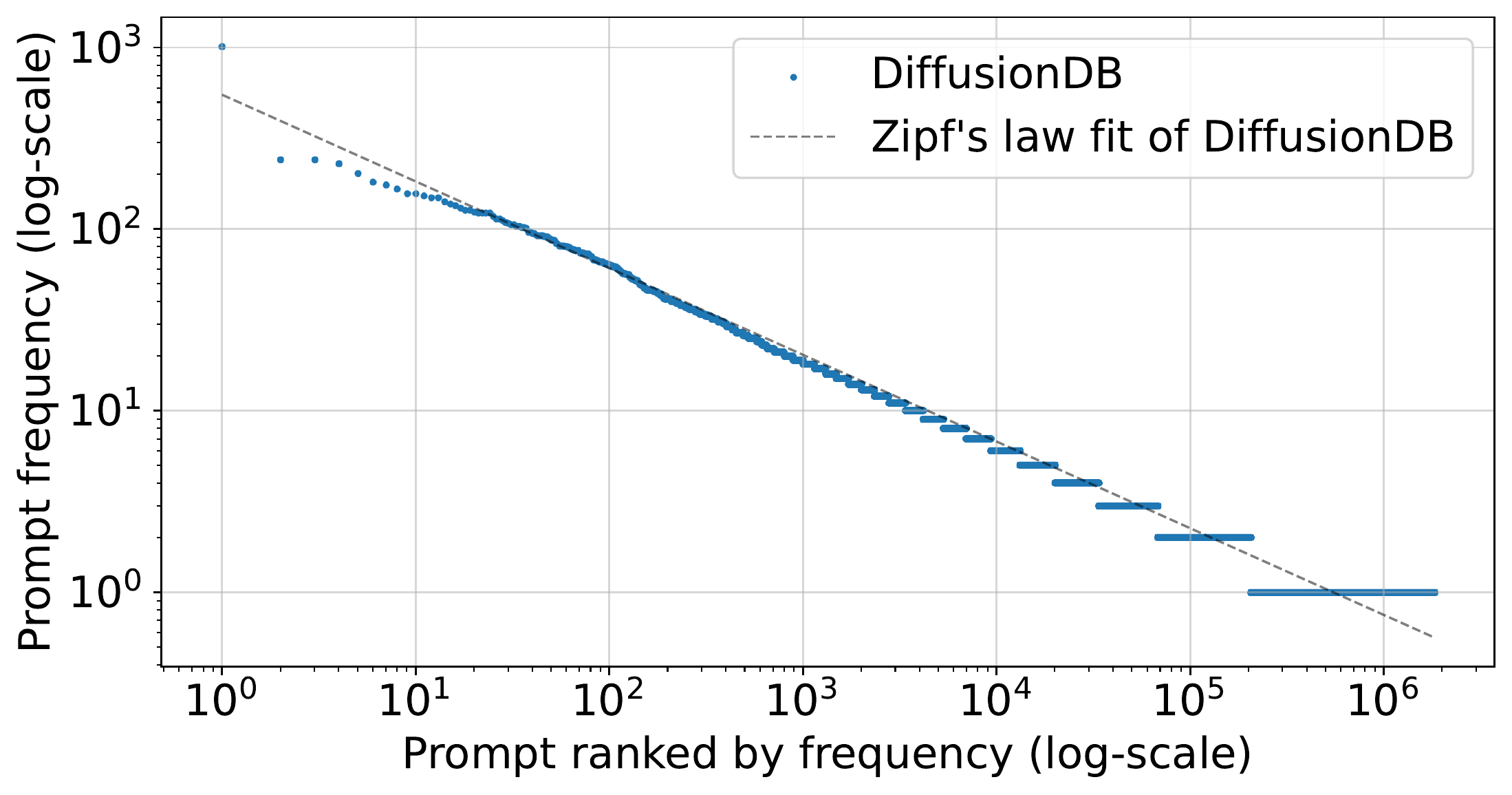}
    \vspace{-15pt}
    \caption{Prompt frequencies of DiffusinoDB plotted in a log-log scale. The distribution follows Zipf's law.
    }
\label{fig:prompt-freq}
\end{figure}

We also examine the distribution of prompt frequencies. From Figure \ref{fig:prompt-freq}, we find the prompt frequency distribution of the larger dataset, DiffusionDB, does follow Zipf's law (except for the very top-ranked prompts), similar to the queries of Web and vertical search engines \cite{silverstein1999analysis,xie2002locality,yang2011query}. The most frequently used prompts are listed in Table \ref{tab:prompt-freq}. 
Interestingly, many of the top-ranked prompts are 
\begin{inparaenum}[(1)]
    \item lengthy and
    \item only used by a few users.
\end{inparaenum}
This indicates that although the prompt frequency distributes are similar to that of Web search, the mechanism underneath may be different (shorter Web queries are more frequent and shared by more users \cite{silverstein1999analysis}). 

\begin{table}[t]
\caption{Most frequent prompts in DiffusionDB.  \#Users indicates the number of users who have used this prompt. }
\vspace{-5pt}
\label{tab:prompt-freq}
\resizebox{\columnwidth}{!}{
\begin{tabular}{r|lrr}
\toprule
\textbf{Rank} & \textbf{Prompt}                         & \hspace{-1cm}\textbf{Freq.} &\hspace{-0.1cm}\textbf{\#Users} \\
\midrule
1  & painful pleasures by lynda benglis, octane render, colorful, 4k, 8k           & 1010          & 1                     \\
2  & cinematic bust portrait of psychedelic robot from left, head and chest ...    & 240           & 2                  \\
3  & divine chaos engine by karol bak, jean delville, william blake, gustav ...    & 240           &  7                   \\
4  & divine chaos engine by karol bak and vincent van gogh                         & 228           &  1                   \\
5  & soft greek sculpture   of intertwined bodies painted by james jean ...        & 202           &  2                  \\
6  & detailed realistic beautiful young medieval queen face portrait ...           & 202           &  1                 \\
7  & animation magic background game design with miss pokemon ...                  & 181           &  2                \\
8  & cat                                                                           & 174           &  69               \\
9  & wrc rally car stylize, art gta 5 cover, official fanart behance hd ...        & 166           &  4             \\
10 & futurism movement hyperrealism 4k detail flat kinetic                         & 157           &  1            \\
11 & a big pile of soft greek sculpture of intertwined bodies painted by ...       & 156           &  1           \\
12 & test                                                                          & 152           &  86          \\
13 & dream                                                                         & 149           &  133         \\
14 & realistic detailed face portrait of a beautiful futuristic viking warrior ... & 149           &  2       \\
15 & spritesheet game asset vector art, smooth style beeple, by thomas ...         & 141           &  3    \\
$^*$16 &                                                                               & 137           &   50   \\
17 & abstract 3d female portrait age five by james jean and jason chan, ...        & 134           &   1  \\
18 & symmetry!! egyptian prince of technology, solid cube of light, ...            & 130           &   1 \\
19 & retrofuturistic portrait of a woman in astronaut helmet, smooth ...           & 127           &   1                  \\
20 & astronaut holding a flag in an underwater desert. a submarine is ...  & 127                   &   1       \\
\bottomrule
\end{tabular}}
\vspace{2pt}
\scriptsize $^*$ Row 16 is an empty prompt.
\vspace{-5pt}
\end{table}

\subsubsection{\textbf{Text-to-image generation prompts tend to be longer. }} 
\label{sec:prompt-length}

We report the key statistics of prompt length (\ie, the number of terms in a prompt) in Table \ref{tab:prompt-length-stats}. The average length of prompts for text-to-image generation (27.16 for Midjourney and 30.34 for DiffusionDB) and the median length (20 for Midjourney and 26 for DiffusionDB) are significantly longer than the lengths of Web search queries, where the mean is around 2.35 and the median is about 2 terms \cite{jansen1998real, silverstein1999analysis}). 
Interestingly, similar observations are reported in vertical search engines such as electronic health records (EHR) search engines, where the queries are also significantly longer than Web search queries (the average length is 5.0) \cite{yang2011query}, likely due to the highly specialized and complex nature of the tasks.  

\begin{table}[t]
\caption{Statistics of prompt lengths. }
\vspace{-5pt}
\label{tab:prompt-length-stats}
\begin{tabular}{l|rrr}
\toprule
\textbf{Dataset}               & \textbf{Midjourney} & \textbf{DiffusionDB} & \textbf{SAC} \\
\midrule
\textbf{Avg. \#terms}    & 27.16	& 30.34	& 17.53	\\
\textbf{Std. \#terms}  & 24.11	& 21.25	& 11.27	\\
\textbf{Median \#terms} & 20	& 26	& 15 	\\
\textbf{Max \#terms}     & 426	& 540	& 62\\\bottomrule
\end{tabular}
\end{table}

\paragraph{Bundled queries}

\begin{figure}[ht]
    \centering
    \includegraphics[width=\linewidth]{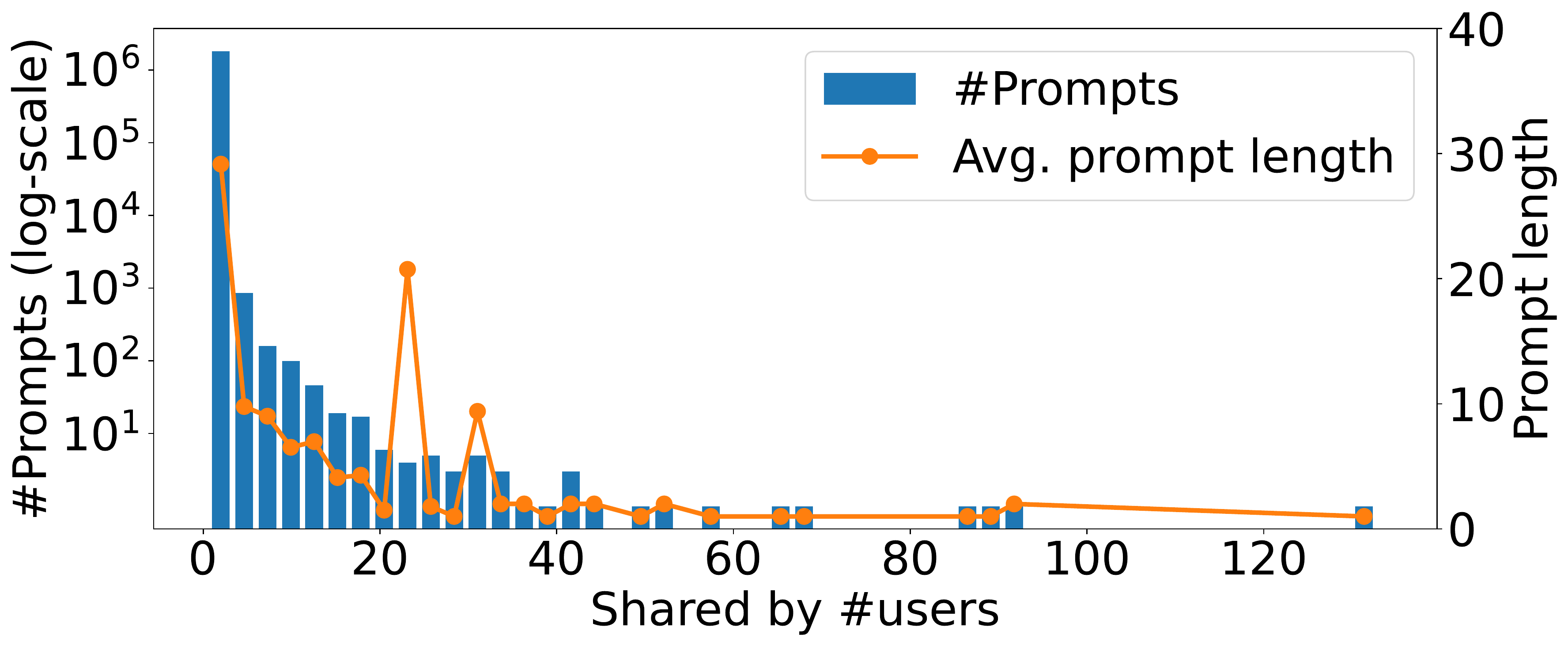}
    \vspace{-15pt}
    \caption{Prompts shared across users in DiffusionDB. 
    The orange line plots the average prompt length in the blue bins. 
    }
\label{fig:prompt-shared}
\end{figure}

When queries are more complex and harder to compose, an effective practice used in medical search engines is to allow users to \textit{bundle} a long query, save it for reuse, and share it with others. In the context of EHR search, \emph{bundled} queries are significantly longer (with 58.9 terms on average, compared to 1.7 terms in user typed-in queries) \cite{zheng2011collaborative,yang2011query}.
Bundled queries tend to have higher quality, and once shared, are more likely to be adopted by other users \cite{zheng2011collaborative}. 
Table~\ref{tab:prompt-freq} seems to suggest the same opportunity, as certain well-composed lengthy queries are revisited many times by their users. These prompts could be saved as ``bundles'' and potentially shared with other users. To illustrate the potential, 
we calculate the prompts used by multiple users and plot the distribution in 
Figure \ref{fig:prompt-shared}. We find a total of 16,950 unique prompts (0.94\% of all unique prompts) have been used across users, 782 have been used by five or more users, and 182 have been shared by 10 or more users. 
The result suggests that text-to-image generation users have already started to share \emph{bundled} prompts spontaneously, even though this functionality has not been provided by the system. Compared to vertical search engines that provide bundle-sharing features, the proportion of \emph{bundled} prompts is still relatively small (compared with 19.3\% for an EHR search engine \cite{yang2011query}), indicating a huge opportunity for bundling and sharing prompts.  

\begin{table}[t]
\caption{Prompts shared by the largest numbers of users in DiffusionDB. 
Only prompts longer than five terms are reported below row 10. }
\vspace{-5pt}
\label{tab:prompt-shared}
\resizebox{\columnwidth}{!}{
\begin{tabular}{r|lr}
\toprule
\textbf{Rank}  & \textbf{Prompt}                         & \hspace{-1.5cm}\textbf{\#Users} \\
\midrule
1  & dream                                                            & 133                                  \\
2  & stable diffusion                                                 & 91                                   \\
3  & help                                                             & 89                                   \\
4  & test                                                             & 86                                   \\
5  & cat                                                              & 69                                   \\
6  & nothing                                                          & 66                                   \\
7  & god                                                              & 58                                   \\
8  & the backrooms                                                    & 53                                   \\
$^*$9  &                                                                  & 50                                   \\
10 & among us                                                         & 44                                   \\
19 & a man standing on top of a bridge over a city, cyberpunk art ... & 32                                   \\
20 & mar - a - lago fbi raid lego set                                 & 32                                   \\
34 & an armchair in the shape of an avocado                           & 23                                   \\
35 & a giant luxury cruiseliner spaceship, shaped like a yacht, ...   & 23                                   \\
42 & a portrait photo of a kangaroo wearing an orange hoodie and ...  & 19                                   \\
45 & anakin skywalker vacuuming the beach to remove sand              & 19                                   \\
48 & emma watson as an avocado chair                                  & 18                                   \\
64 & milkyway in a glass bottle, 4k, unreal engine, octane render     & 16                                  \\
\bottomrule
\end{tabular}}
\vspace{2pt}
\scriptsize $^*$ Row 9 is an empty prompt.
\vspace{-10pt}
\end{table}

\subsubsection{\textbf{Text-to-image generation sessions contain more prompts.} }
\label{sec:session}
A session is defined as a sequence of queries made by the same user within a short time frame in Web search \cite{yang2011query}, which often corresponds to an atomic mission for a user to achieve a single information need \cite{silverstein1999analysis,jones2008beyond}. 
Analyzing sessions is critical in query log analysis because a session provides insights about how a user modifies the queries to fulfill the information need \cite{silverstein1999analysis, jones2008beyond}. 

Following the common practice in Web search, we chunk prompts into sessions with a 30-minute timeout \cite{jansen1998real,yang2011query}, meaning any two consecutive prompts that are submitted by the same user within 30 minutes will be considered as in the same session. 

The statistics of sessions are listed in Table \ref{tab:prompt-session-stats}. Similar to prompts, text-to-image generation sessions also tend to be significantly longer than Web search sessions (by the number of prompts in a session). A text-to-image generation session contains 10.25 or 13.71 (Midjourney or DiffusionDB) prompts on average and a median of 4 or 5 (Midjourney or DiffusionDB) prompts; while in Web search, the average session length is around 2.02 and the median is 1 \cite{silverstein1999analysis}. This is again likely due to the complexity of the creation task so the users need to update the prompts multiple times. Indeed, a user tends to change (\emph{add}, \emph{delete}, or \emph{replace}) a median of 3 terms (measured by term-level edit distance) between two consecutive prompts in the same session on Midjourney (5 on DiffusionDB), astonishingly more than how people update Web search queries. Do these updates indicate different types of information needs? 

\begin{table}[htbp]
\caption{Statistics of prompt sessions. Sessions are identified with a 30-minute timeout. Edit distances regarding terms are calculated with consecutive prompts in the same session. }
\vspace{-5pt}
\label{tab:prompt-session-stats}
\begin{tabular}{l|rr}
\toprule
\textbf{Dataset}                    & \textbf{Midjourney} & \textbf{DiffusionDB} \\
\midrule
\textbf{\#Sessions}                 & 14,232                    & 161,001                     \\
\textbf{Avg. \#sessions/user}   & 8.52                & 15.51                \\
\textbf{Median \#sessions/user} & 2                   & 9                    \\
\midrule
\textbf{Avg. \#prompts/session}        & 10.19               & 13.71                \\
\textbf{Median \#prompts/session}      & 4                   & 5                    \\
\midrule
\textbf{Avg. edit distance}         & 8.53                & 9.42                 \\
\textbf{Median edit distance}       & 3                   & 5                    \\
\bottomrule
\end{tabular}
\vspace{-5pt}
\end{table}

\subsubsection{\textbf{A new categorization of information needs.}}
\label{sec:prompt-categorization}

Web search queries are typically distinguished into three categories: 
\begin{inparaenum}[(1)]
    \item \emph{navigational queries},
    \item \emph{informational queries}, and
    \item \emph{transactional queries} 
\end{inparaenum}
\cite{broder2002taxonomy}. 
Should text-to-image prompts be categorized in the same way?
\textcolor{mygreen}{
Or do prompts express new categories of information needs?
}

\paragraph{Navigational prompts}

The most frequent queries in Web search are often \emph{navigational}, where users simply use a query to lead them to a particular, known Website (\eg, ``Facebook'' or ``YouTube''). 
In text-to-image generation, as the generation model often returns different images given the same text prompt due to randomization, the information need of ``navigating'' to a known image is rare. Indeed, the queries used by the most number of users (Figure \ref{fig:prompt-shared}) are generally not tied to a particular image. Even though the shorter queries on the top look somewhat similar to ``Facebook'' or ``Youtube'', are rather ambiguous and more like testing the system. 


\paragraph{Informational prompts}

Most other text-to-image prompts can be compared to \emph{informational} queries in Web search, which aim to acquire certain information that is expected to present on one or more Web pages \cite{broder2002taxonomy}. The difference is that informational prompts aim to \emph{synthesize} (rather than \emph{retrieve}) an image, which is expected to exist in the \emph{latent} representation space of images. 
Most prompts fall into this category, similar to the case in Web search \cite{broder2002taxonomy}. 

\paragraph{Transactional prompts}

\emph{Transactional} queries are those intended of performing certain Web-related activities \cite{broder2002taxonomy}, such as completing a transaction (\eg, to book a flight or to make a purchase). 
One could superficially categorize all prompts into transactional, as they are all intended to conduct the activities of ``generating images''.  Zooming into this superficial categorization, we could identify prompts that refer to specific and recurring tasks, such as ``3D rendering'', ``post-processing'', ``global illumination'', and ``movie poster'' (see more examples in Section~\ref{sec:applications}). These tasks may be considered transactional in the context of text-to-image generation.  

\paragraph{Exploratory prompts}

Beyond the above categories corresponding to the three basic types of Web search queries, we discover a new type of information needs in prompts, namely the \emph{exploratory} prompts for text-to-image generation. Comparing to an \emph{informational} prompt that aims to generate a specific piece of (hypothetical) image, an \emph{exploratory} prompt often describes a vague or uncertain information need (or image generation requirements) that intentionally leads to multiple possible answers. The user intends to explore different possibilities, leveraging either the randomness of the model or the flexibility of terms used in a prompt session. 

Indeed, rather than clearly specifying the requirements and constraints and gradually refining the requirements in a session, in \emph{exploratory} prompts or sessions, the users tend to play with alternative terms of the same category (\eg, different colors or animals, or \textit{sibling} terms) to explore how the generation results could be different or could cover a broader search space. Based on the session analysis, we count the most frequent term replacements in Table \ref{tab:term-replacement}. In this table, we find 33 replacements that show \emph{exploratory} patterns, such as (``man'', ``woman''), (``asian'', ``white''), (``dog'', ``cat''), (``red'', ``blue''), and (``16:9'', ``9:16''). 

On the contrary, in \emph{non-exploratory} sessions, replacing a term with its synonyms or hyponyms, or more specific concepts are more common, which refines the search space (rather than exploring the generation space). In the table, we find a few such replacements: (``steampunk'', ``cyberpunk''), (``deco'', ``nouveau''), (``crown'', ``throne''). There are also examples that replace terms with the correct spelling or replace punctuations to refine: (``aphrodesiac'', ``aphrodisiac''), (``with'', ``,''), (``,'', ``and'') and (``,'', ``.'').

\definecolor{cellgreen}{HTML}{C6EFCE}
\begin{table}[t]
\caption{Most frequent term replacements. This table only considers consecutive prompts from the same session where exactly one term is been replaced. \colorbox{cellgreen}{Green} highlights replacements that might indicates \emph{exploratory} patterns, while \colorbox{pink}{red} highlights \emph{non-exploratory} replacements. 
}
\vspace{-5pt}
\label{tab:term-replacement}
\resizebox{0.95\columnwidth}{!}{
\begin{tabular}{r|lr|lr}
\toprule
   & \multicolumn{2}{c|}{\textbf{Midjourney}} & \multicolumn{2}{c}{\textbf{DiffusionDB}}      \\
\hhline{~----}
   & \textbf{Replacement}                      & \hspace{-1.5cm}\textbf{Freq.}  & \textbf{Replacement}           & \hspace{-1.5cm}\textbf{Freq.}  \\
\midrule
1	&	\colorbox{cellgreen}{('deco', 'nouveau')}	&	16	&	\colorbox{cellgreen}{('man', 'woman')}	&	216	\\
2	&	\colorbox{cellgreen}{('16:9', '9:16')}	&	15	&	\colorbox{cellgreen}{('woman', 'man')}	&	187	\\
3	&	\colorbox{cellgreen}{('9:16', '16:9')}	&	14	&	\colorbox{white}{('2', '3')}	&	161	\\
4	&	\colorbox{cellgreen}{('2', '1')}	&	8	&	\colorbox{white}{('1', '2')}	&	147	\\
5	&	\colorbox{cellgreen}{('16:9', '4:6')}	&	8	&	\colorbox{white}{('7', '8')}	&	140	\\
6	&	\colorbox{white}{('1', '2')}	&	7	&	\colorbox{white}{('8', '9')}	&	139	\\
7	&	\colorbox{cellgreen}{('3:4', '4:3')}	&	7	&	\colorbox{white}{('6', '7')}	&	135	\\
8	&	\colorbox{white}{('1000', '10000')}	&	7	&	\colorbox{white}{('3', '4')}	&	132	\\
9	&	\colorbox{white}{('artwork', 'parrot')}	&	7	&	\colorbox{cellgreen}{('girl', 'woman')}	&	128	\\
10 &	\colorbox{cellgreen}{('16:9', '1:2')}	&	6	&	\colorbox{cellgreen}{('red', 'blue')}	&	116	\\	
11 &	\colorbox{cellgreen}{('2:3', '3:2')}	&	6	&	\colorbox{white}{('5', '6')}	&	115	\\	
12 &	\colorbox{cellgreen}{('asian', 'white')}	&	6	&	\colorbox{white}{('4', '5')}	&	112	\\	
13 &	\colorbox{cellgreen}{('1', '0.5')}	&	5	&	\colorbox{cellgreen}{('female', 'male')}	&	107	\\
14 &	\colorbox{white}{('320', '384')}	&	5	&	\colorbox{cellgreen}{('male', 'female')}	&	97	\\	
15 &	\colorbox{white}{('0.5', '1')}	&	4	&	\colorbox{cellgreen}{('blue', 'red')}	&	93	\\	
16 &	\colorbox{pink}{('crown', 'throne')}	&	4	&	\colorbox{white}{('0', '1')}	&	89	\\
17 &	\colorbox{cellgreen}{('blue', 'green')}	&	4	&	\colorbox{cellgreen}{('cat', 'dog')}	&	89	\\	
18 &	\colorbox{cellgreen}{('9:16', '4:5')}	&	4	&	\colorbox{cellgreen}{('woman', 'girl')}	&	82	\\	
19 &	\colorbox{cellgreen}{('2:3', '1:2')}	&	4	&	\colorbox{cellgreen}{('dog', 'cat')}	&	79	\\	
20 &	\colorbox{cellgreen}{('-{}-w', '-{}-h')}	&	4	&	\colorbox{cellgreen}{('white', 'black')}	&	72	\\
21 &	\colorbox{cellgreen}{('nouveau', 'deco')}	&	4	&	\colorbox{pink}{('with', ',')}	&	71	\\	
22 &	\colorbox{cellgreen}{('red', 'blue')}	&	4	&	\colorbox{pink}{('steampunk', 'cyberpunk')}	&	71	\\
23 &	\colorbox{cellgreen}{('guy', 'girl')}	&	4	&	\colorbox{cellgreen}{('red', 'green')}	&	70	\\	
24 &	\colorbox{white}{('snake', 'apple')}	&	4	&	\colorbox{pink}{('cyberpunk', 'steampunk')}	&	70	\\	
25 &	\colorbox{cellgreen}{('japanese', 'korean')}	&	4	&	\colorbox{pink}{(',', 'and')}	&	69	\\	
26 &	\colorbox{cellgreen}{('16:8', '8:11')}	&	4	&	\colorbox{pink}{('painting', 'portrait')}	&	68	\\	
27 &	\colorbox{pink}{('insect', 'ladybug')}	&	4	&	\colorbox{pink}{(',', '.')}	&	68	\\	
28 &	\colorbox{cellgreen}{('-{}-hd', '-{}-vibe')}	&	3	&	\colorbox{pink}{('portrait', 'painting')}	&	68	\\	
29 &	\colorbox{pink}{('aphrodesiac', 'aphrodisiac')}	&	3	&	\colorbox{cellgreen}{('girl', 'boy')}	&	64	\\	
30 &	\colorbox{white}{('0.5', '2')}	&	3	&	\colorbox{cellgreen}{('green', 'blue')}	&	63	\\	
\bottomrule
\end{tabular}}
\vspace{-10pt}
\end{table}

Another indication of \emph{exploratory} behavior is the repeated use of prompts. For example, among the top prompts in Table \ref{tab:prompt-freq} (except those for testing purposes), each of them is repeatedly used by the same user more than 100 times. This might be because the user is exploring different generation results with the same prompt, leveraging the randomness of the generative model. 

\subsection{How are the Information Needs Satisfied?}
\label{sec:prompts-ratings}

Prompts are typically crafted to \textcolor{mygreen}{meet certain information needs by generating} satisfactory images. In this subsection, we examine how prompts can fulfill this goal. 
With the rating annotations in the SAC dataset (the average rating is 5.53, and the median is 6), 
we calculate the correlation between ratings and other variables like prompt lengths and term frequencies. 

\subsubsection{\textbf{Longer prompts tend to be higher rated. }} 

\begin{figure}[htbp]
  \centering
  \includegraphics[width=\linewidth]{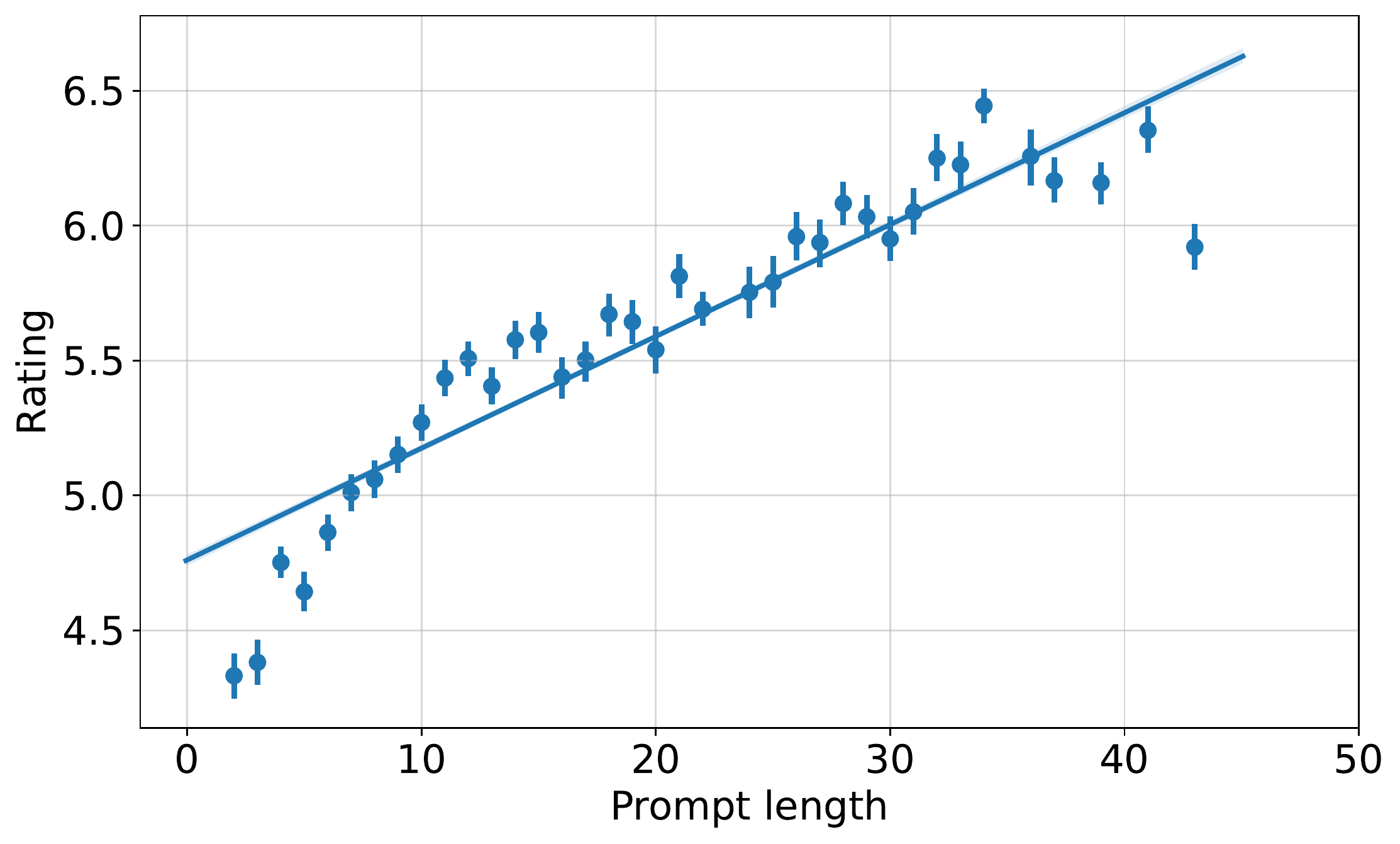}
  \vspace{-15pt}
  \caption{Prompt length is positively correlated with ratings. The Pearson correlation coefficient is 0.197.  }
  \label{fig:ratings-vs-lengths}
\end{figure}

We plot how the ratings of generated images correlate with prompt lengths in Figure \ref{fig:ratings-vs-lengths}, where we find a positive correlation with the Pearson coefficient at 0.197. This means longer prompts tend to produce images of higher quality. This provides another perspective to understand the large lengths of prompts and prompt sessions and another motivation to bundle and share long prompts. 

\subsubsection{\textbf{The choice of words matters.} }

We also investigate how the choice of words influences the performance of image generation. We collect all the prompts that contain a particular term and calculate the average rating. 
Terms with the highest and lowest average ratings are listed in Table \ref{tab:terms-highest-lowest-ratings} in the appendix. 
We find most high-rating terms are artist names, which provide clear constraints on the styles of images.
In contrast, terms with low ratings are much vaguer and more abstract and might indicate an \emph{exploratory} behavior. More efforts needed to be done to handle \emph{exploratory} prompts and to encourage the users to refine their needs. 

\subsection{How are Users' Information Needs Covered by Image Captions?}
\label{sec:prompt-train-diff}

Current text-to-image generation models are generally trained with large-scale image-text datasets, where the paired text usually come from image captions. To figure out how these training sets match the actual users' information needs, we compare the prompts with image captions in the open domain. In particular, we consider LAION-400M \cite{schuhmann2021laion} as one of the main sources of text-to-image training data since both LDMs and the Stable Diffusion model employ this dataset. Text in LAION-400M are extracted from the captions of the images collected from the Common Crawl, so they are supposed to convey the subject, form, and intent of the images.  
We randomly sample 1M texts from LAION-400M and compare them with user-input prompts. 
We obtain the following finding. 

\paragraph{\textbf{Term usages are different between user-input prompts and image captions in open datasets}} 

We construct a vocabulary based on LAION-400M and calculate the vocabulary coverage of three prompt datasets (\ie, to what proportion of the user-input terms is covered by the LAION vocabulary). The coverage is 25.94\% for Midjourney, 43.17\% for DiffusionDB, and 80.56\% for SAC.
The coverage is relatively high on SAC as this dataset is relatively clean. In comparison, the Midjourney and DifussionDB datasets directly collect prompts from Discord channels of Midjourney and Stable Diffusion, and over half of the terms are not covered in the LAION dataset. 
We also analyzed their embeddings and find that user-input prompts and image captions from the LAION dataset cover very different regions in the latent space (Figure \ref{fig:prompt-emb} in the appendix). 

\section{Implications}
\label{sec:challenges}

Our analysis presents unique characteristics of user-input prompts, which helps us better understand the limitations and opportunities of text-to-image generation systems and AI-facilitated creativity on the Web. Below we discuss a few concrete and actionable possibilities for improving the generation systems and enhancing creativity. 

\paragraph{\textbf{Building art creativity glossaries}}

As we discussed in Sec. \ref{sec:art-components}, a text-to-image prompt could be decomposed into three aspects: \emph{subject} (``what''), \emph{form} (``how''), and \emph{intent} (``why'', or \emph{content} as in classical Art literature). 
If we can identify and analyze these specific elements in prompts, we may be able to better decipher users' information needs. 

However, to the best of our knowledge, there is no existing tool that is able to extract the \emph{subject}, \emph{form}, and \emph{intent} from text prompts. Besides, although users have spontaneously collected terms that describe the \emph{form} and \emph{subject}\footnote{Prompt book for data lovers II: \url{https://docs.google.com/presentation/d/1V8d6TIlKqB1j5xPFH7cCmgKOV_fMs4Cb4dwgjD5GIsg}, retrieved on 3/14/2023.},
there is no high-quality and comprehensive glossary in the literature that contains terms about these three basic components of art, or something like the Unified Medical Language System (UMLS) for biomedical and health domains \cite{bodenreider2004unified}. Constructing such tools or glossaries is difficult and will highly rely on the domain knowledge, because:
\begin{inparaenum}[(1)]
    \item These three components of art are often intertwined and inseparable in a piece of work \cite{ocvirk1968art}, meaning a term would have tendencies to fall into any categories of these three. For example, in \emph{Process Art}, the \emph{form} and \emph{content} seem to be the same thing \cite{ocvirk1968art}. 
    \item Terminologies about art are consistently updated
    because new artists, styles, and art-related sites keep popping out. 
\end{inparaenum}
We call for the joint effort of the art and the Web communities to build such vocabularies and tools. 

\paragraph{\textbf{Bundling and sharing prompts}}

Sec. \ref{sec:prompt-length} analyzes the lengths of text-to-image prompts, where we find an inadequate use of \emph{bundled} prompts compared with other vertical search engines (\eg, EHR search engines). 
Since the prompts are generally much longer than Web search queries, and the information needs are also more complex, it is highly likely that \emph{bundled} prompts can help the users to craft their prompts more effectively and efficiently. 
Though there are already prompt search websites like Lexica\footnote{Lexica: \url{https://lexica.art/}, retrieved on 3/14/2023.}, PromptHero\footnote{PromptHero: \url{https://prompthero.com/}, retrieved on 3/14/2023. } and PromptBase\footnote{PromptBase: \url{https://promptbase.com/}, retrieved on 3/14/2023. } that provide millions of user-crafted prompts, such \emph{bundled} search features are merely integrated into current text-to-image generation systems. As mentioned earlier, adding features to support bundling and sharing high-quality prompts could bring immediate benefits to text-to-image generation systems. 

\paragraph{\textbf{Personalized generation}}

The analysis in Sec. \ref{sec:session} suggests that the session lengths in text-to-image generation are also significantly larger than the session lengths in Web search, indicating the great opportunity for a personalized generation. 
Currently, the \emph{session-based} generation features are mostly built upon image initialization of diffusion models, \ie, using the output from the previous generation as the starting point of diffusion sampling. 
Compared with other \emph{session-based} AI systems like ChatGPT \cite{chatgpt}, these \emph{session-based} features still seem preliminary and take little consideration about personalized generation. Meanwhile, the explicit descriptions of \emph{forms} and \emph{intent} in prompts also indicate opportunities to customize the generation models for these constraints (and the potential applications as listed in Section~\ref{sec:applications}).   

\paragraph{\textbf{Handling exploratory prompts and sessions}}

In Sec. \ref{sec:prompt-categorization} we identify a new type of prompt in addition to the three typical categories of query in Web search (\ie, \emph{navigational}, \emph{informational}, and \emph{transactional} queries), namely the \emph{exploratory} prompts. 
To encourage the \emph{exploratory} generation of images, reliable and informative \emph{exploration measures} will be much needed. In other machine innovation areas, like AI for molecular generation, efforts have been made on discussing the measurement of coverage and exploration of spaces \cite{xie2022much,xie2023how}, but for text-to-image generation, such discussions are still rare. How to encourage the models to explore a larger space, generate novel and diverse images, and recommend exploratory prompts to users are all promising yet challenging directions. 

\paragraph{\textbf{Improving generation models with prompt logs.}} Finally, the gap between the image captions in open datasets and the user-input prompts (Sec. \ref{sec:prompt-train-diff}) indicates that it is desirable to improve model training directly using the prompt logs. Following the common practice in Web search engines, one may leverage both explicit and implicit feedback from the prompt logs (such as the ratings or certain behavioral patterns or modifications in the prompts) as additional signals to update the generation models. 

Although we focus our analysis on text-to-image generation, the analogy to Web search and some of the above implications also apply to other domains of AI-generated content (AIGC), such as AI chatbots (\eg, ChatGPT).  

\section{Conclusion}

We take an initial step to investigate the information needs of text-to-image generation through a comprehensive and large-scale analysis of user-input prompts (analogous to Web search queries) in multiple popular systems.
The results suggest that 
\begin{inparaenum}[(1)]
    \item text-to-image prompts are typically structured with terms that describe the \emph{subject}, \emph{form}, and \emph{intent};
    \item text-to-image prompts are sufficiently different from Web search queries. Our findings include the significantly lengthier prompts and sessions, the lack of \emph{navigational} prompts, the new perspective of \emph{transactional} prompts, and the prevalence of \emph{exploratory} prompts;
    \item image generation quality is correlated with the length of the prompt as well as the usage of terms; and 
    \item there is a considerable gap between the user-input prompts and the image captions used to train the models. 
\end{inparaenum}
Based on these findings, we present actionable insights to improve text-to-image generation systems. 
We anticipate our study could help the text-to-image generation community to better understand and facilitate creativity on the Web. 

\newpage
\bibliographystyle{ACM-Reference-Format}
\bibliography{sample-base}


\begin{thebibliography}{46}


\ifx \showCODEN    \undefined \def \showCODEN     #1{\unskip}     \fi
\ifx \showDOI      \undefined \def \showDOI       #1{#1}\fi
\ifx \showISBNx    \undefined \def \showISBNx     #1{\unskip}     \fi
\ifx \showISBNxiii \undefined \def \showISBNxiii  #1{\unskip}     \fi
\ifx \showISSN     \undefined \def \showISSN      #1{\unskip}     \fi
\ifx \showLCCN     \undefined \def \showLCCN      #1{\unskip}     \fi
\ifx \shownote     \undefined \def \shownote      #1{#1}          \fi
\ifx \showarticletitle \undefined \def \showarticletitle #1{#1}   \fi
\ifx \showURL      \undefined \def \showURL       {\relax}        \fi
\providecommand\bibfield[2]{#2}
\providecommand\bibinfo[2]{#2}
\providecommand\natexlab[1]{#1}
\providecommand\showeprint[2][]{arXiv:#2}

\bibitem[Agresti(2012)]%
        {agresti2012categorical}
\bibfield{author}{\bibinfo{person}{Alan Agresti}.}
  \bibinfo{year}{2012}\natexlab{}.
\newblock \bibinfo{booktitle}{\emph{Categorical data analysis}}.
  Vol.~\bibinfo{volume}{792}.
\newblock \bibinfo{publisher}{John Wiley \& Sons}.
\newblock


\bibitem[Bodenreider(2004)]%
        {bodenreider2004unified}
\bibfield{author}{\bibinfo{person}{Olivier Bodenreider}.}
  \bibinfo{year}{2004}\natexlab{}.
\newblock \showarticletitle{The unified medical language system (UMLS):
  integrating biomedical terminology}.
\newblock \bibinfo{journal}{\emph{Nucleic acids research}}
  \bibinfo{volume}{32}, \bibinfo{number}{suppl\_1} (\bibinfo{year}{2004}),
  \bibinfo{pages}{D267--D270}.
\newblock


\bibitem[Broder(2002)]%
        {broder2002taxonomy}
\bibfield{author}{\bibinfo{person}{Andrei Broder}.}
  \bibinfo{year}{2002}\natexlab{}.
\newblock \showarticletitle{A taxonomy of web search}. In
  \bibinfo{booktitle}{\emph{ACM Sigir forum}}, Vol.~\bibinfo{volume}{36}. ACM
  New York, NY, USA, \bibinfo{pages}{3--10}.
\newblock


\bibitem[Brown et~al\mbox{.}(2020)]%
        {brown2020language}
\bibfield{author}{\bibinfo{person}{Tom Brown}, \bibinfo{person}{Benjamin Mann},
  \bibinfo{person}{Nick Ryder}, \bibinfo{person}{Melanie Subbiah},
  \bibinfo{person}{Jared~D Kaplan}, \bibinfo{person}{Prafulla Dhariwal},
  \bibinfo{person}{Arvind Neelakantan}, \bibinfo{person}{Pranav Shyam},
  \bibinfo{person}{Girish Sastry}, \bibinfo{person}{Amanda Askell},
  {et~al\mbox{.}}} \bibinfo{year}{2020}\natexlab{}.
\newblock \showarticletitle{Language models are few-shot learners}.
\newblock \bibinfo{journal}{\emph{Advances in neural information processing
  systems}}  \bibinfo{volume}{33} (\bibinfo{year}{2020}),
  \bibinfo{pages}{1877--1901}.
\newblock


\bibitem[Cho et~al\mbox{.}(2020)]%
        {Cho2020XLXMERT}
\bibfield{author}{\bibinfo{person}{Jaemin Cho}, \bibinfo{person}{Jiasen Lu},
  \bibinfo{person}{Dustin Schwenk}, \bibinfo{person}{Hannaneh Hajishirzi},
  {and} \bibinfo{person}{Aniruddha Kembhavi}.} \bibinfo{year}{2020}\natexlab{}.
\newblock \showarticletitle{X-LXMERT: Paint, Caption and Answer Questions with
  Multi-Modal Transformers}. In \bibinfo{booktitle}{\emph{EMNLP}}.
\newblock


\bibitem[Chowdhery et~al\mbox{.}(2022)]%
        {chowdhery2022palm}
\bibfield{author}{\bibinfo{person}{Aakanksha Chowdhery},
  \bibinfo{person}{Sharan Narang}, \bibinfo{person}{Jacob Devlin},
  \bibinfo{person}{Maarten Bosma}, \bibinfo{person}{Gaurav Mishra},
  \bibinfo{person}{Adam Roberts}, \bibinfo{person}{Paul Barham},
  \bibinfo{person}{Hyung~Won Chung}, \bibinfo{person}{Charles Sutton},
  \bibinfo{person}{Sebastian Gehrmann}, {et~al\mbox{.}}}
  \bibinfo{year}{2022}\natexlab{}.
\newblock \showarticletitle{Palm: Scaling language modeling with pathways}.
\newblock \bibinfo{journal}{\emph{arXiv preprint arXiv:2204.02311}}
  (\bibinfo{year}{2022}).
\newblock


\bibitem[Christiano et~al\mbox{.}(2017)]%
        {christiano2017deep}
\bibfield{author}{\bibinfo{person}{Paul~F Christiano}, \bibinfo{person}{Jan
  Leike}, \bibinfo{person}{Tom Brown}, \bibinfo{person}{Miljan Martic},
  \bibinfo{person}{Shane Legg}, {and} \bibinfo{person}{Dario Amodei}.}
  \bibinfo{year}{2017}\natexlab{}.
\newblock \showarticletitle{Deep reinforcement learning from human
  preferences}.
\newblock \bibinfo{journal}{\emph{Advances in neural information processing
  systems}}  \bibinfo{volume}{30} (\bibinfo{year}{2017}).
\newblock


\bibitem[Clauset et~al\mbox{.}(2009)]%
        {clauset2009power}
\bibfield{author}{\bibinfo{person}{Aaron Clauset},
  \bibinfo{person}{Cosma~Rohilla Shalizi}, {and} \bibinfo{person}{Mark~EJ
  Newman}.} \bibinfo{year}{2009}\natexlab{}.
\newblock \showarticletitle{Power-law distributions in empirical data}.
\newblock \bibinfo{journal}{\emph{SIAM review}} \bibinfo{volume}{51},
  \bibinfo{number}{4} (\bibinfo{year}{2009}), \bibinfo{pages}{661--703}.
\newblock


\bibitem[Davenport and Mittal(2022)]%
        {davenport_mittal_2022}
\bibfield{author}{\bibinfo{person}{Thomas~H. Davenport} {and}
  \bibinfo{person}{Nitin Mittal}.} \bibinfo{year}{2022}\natexlab{}.
\newblock \bibinfo{title}{How generative AI is changing creative work}.
\newblock
\newblock
\urldef\tempurl%
\url{https://hbr.org/2022/11/how-generative-ai-is-changing-creative-work}
\showURL{%
\tempurl}
\newblock
\shownote{Retrieved on 3/15/2023.}.


\bibitem[Devlin et~al\mbox{.}(2019)]%
        {devlin2018bert}
\bibfield{author}{\bibinfo{person}{Jacob Devlin}, \bibinfo{person}{Ming-Wei
  Chang}, \bibinfo{person}{Kenton Lee}, {and} \bibinfo{person}{Kristina
  Toutanova}.} \bibinfo{year}{2019}\natexlab{}.
\newblock \showarticletitle{{BERT}: Pre-training of Deep Bidirectional
  Transformers for Language Understanding}. In
  \bibinfo{booktitle}{\emph{Proceedings of the 2019 Conference of the North
  {A}merican Chapter of the Association for Computational Linguistics: Human
  Language Technologies, Volume 1 (Long and Short Papers)}}.
  \bibinfo{publisher}{Association for Computational Linguistics},
  \bibinfo{address}{Minneapolis, Minnesota}, \bibinfo{pages}{4171--4186}.
\newblock
\urldef\tempurl%
\url{https://doi.org/10.18653/v1/N19-1423}
\showDOI{\tempurl}


\bibitem[Goodfellow et~al\mbox{.}(2014)]%
        {NIPS2014_5ca3e9b1}
\bibfield{author}{\bibinfo{person}{Ian Goodfellow}, \bibinfo{person}{Jean
  Pouget-Abadie}, \bibinfo{person}{Mehdi Mirza}, \bibinfo{person}{Bing Xu},
  \bibinfo{person}{David Warde-Farley}, \bibinfo{person}{Sherjil Ozair},
  \bibinfo{person}{Aaron Courville}, {and} \bibinfo{person}{Yoshua Bengio}.}
  \bibinfo{year}{2014}\natexlab{}.
\newblock \showarticletitle{Generative Adversarial Nets}. In
  \bibinfo{booktitle}{\emph{Advances in Neural Information Processing
  Systems}}, \bibfield{editor}{\bibinfo{person}{Z.~Ghahramani},
  \bibinfo{person}{M.~Welling}, \bibinfo{person}{C.~Cortes},
  \bibinfo{person}{N.~Lawrence}, {and} \bibinfo{person}{K.Q. Weinberger}}
  (Eds.), Vol.~\bibinfo{volume}{27}. \bibinfo{publisher}{Curran Associates,
  Inc.}
\newblock


\bibitem[Herskovic et~al\mbox{.}(2007)]%
        {herskovic2007day}
\bibfield{author}{\bibinfo{person}{Jorge~R Herskovic}, \bibinfo{person}{Len~Y
  Tanaka}, \bibinfo{person}{William Hersh}, {and} \bibinfo{person}{Elmer~V
  Bernstam}.} \bibinfo{year}{2007}\natexlab{}.
\newblock \showarticletitle{A day in the life of PubMed: analysis of a typical
  day's query log}.
\newblock \bibinfo{journal}{\emph{Journal of the American Medical Informatics
  Association}} \bibinfo{volume}{14}, \bibinfo{number}{2}
  (\bibinfo{year}{2007}), \bibinfo{pages}{212--220}.
\newblock


\bibitem[Ho et~al\mbox{.}(2020)]%
        {ho2020denoising}
\bibfield{author}{\bibinfo{person}{Jonathan Ho}, \bibinfo{person}{Ajay Jain},
  {and} \bibinfo{person}{Pieter Abbeel}.} \bibinfo{year}{2020}\natexlab{}.
\newblock \showarticletitle{Denoising diffusion probabilistic models}.
\newblock \bibinfo{journal}{\emph{Advances in Neural Information Processing
  Systems}}  \bibinfo{volume}{33} (\bibinfo{year}{2020}),
  \bibinfo{pages}{6840--6851}.
\newblock


\bibitem[Jansen et~al\mbox{.}(1998)]%
        {jansen1998real}
\bibfield{author}{\bibinfo{person}{Bernard~J Jansen}, \bibinfo{person}{Amanda
  Spink}, \bibinfo{person}{Judy Bateman}, {and} \bibinfo{person}{Tefko
  Saracevic}.} \bibinfo{year}{1998}\natexlab{}.
\newblock \showarticletitle{Real life information retrieval: A study of user
  queries on the web}. In \bibinfo{booktitle}{\emph{ACM Sigir Forum}},
  Vol.~\bibinfo{volume}{32}. ACM New York, NY, USA, \bibinfo{pages}{5--17}.
\newblock


\bibitem[Jones and Klinkner(2008)]%
        {jones2008beyond}
\bibfield{author}{\bibinfo{person}{Rosie Jones} {and}
  \bibinfo{person}{Kristina~Lisa Klinkner}.} \bibinfo{year}{2008}\natexlab{}.
\newblock \showarticletitle{Beyond the session timeout: automatic hierarchical
  segmentation of search topics in query logs}. In
  \bibinfo{booktitle}{\emph{Proceedings of the 17th ACM conference on
  Information and knowledge management}}. \bibinfo{pages}{699--708}.
\newblock


\bibitem[Kingma and Welling(2013)]%
        {kingma2013auto}
\bibfield{author}{\bibinfo{person}{Diederik~P Kingma} {and}
  \bibinfo{person}{Max Welling}.} \bibinfo{year}{2013}\natexlab{}.
\newblock \showarticletitle{Auto-encoding variational bayes}.
\newblock \bibinfo{journal}{\emph{arXiv preprint arXiv:1312.6114}}
  (\bibinfo{year}{2013}).
\newblock


\bibitem[Liu and Chilton(2022)]%
        {liu2022design}
\bibfield{author}{\bibinfo{person}{Vivian Liu} {and} \bibinfo{person}{Lydia~B
  Chilton}.} \bibinfo{year}{2022}\natexlab{}.
\newblock \showarticletitle{Design guidelines for prompt engineering
  text-to-image generative models}. In \bibinfo{booktitle}{\emph{Proceedings of
  the 2022 CHI Conference on Human Factors in Computing Systems}}.
  \bibinfo{pages}{1--23}.
\newblock


\bibitem[Mansimov et~al\mbox{.}(2016)]%
        {mansimov2015generating}
\bibfield{author}{\bibinfo{person}{Elman Mansimov}, \bibinfo{person}{Emilio
  Parisotto}, \bibinfo{person}{Jimmy Ba}, {and} \bibinfo{person}{Ruslan
  Salakhutdinov}.} \bibinfo{year}{2016}\natexlab{}.
\newblock \showarticletitle{Generating Images from Captions with Attention}. In
  \bibinfo{booktitle}{\emph{ICLR}}.
\newblock


\bibitem[McInnes et~al\mbox{.}(2018)]%
        {mcinnes2018umap}
\bibfield{author}{\bibinfo{person}{Leland McInnes}, \bibinfo{person}{John
  Healy}, {and} \bibinfo{person}{James Melville}.}
  \bibinfo{year}{2018}\natexlab{}.
\newblock \showarticletitle{Umap: Uniform manifold approximation and projection
  for dimension reduction}.
\newblock \bibinfo{journal}{\emph{arXiv preprint arXiv:1802.03426}}
  (\bibinfo{year}{2018}).
\newblock


\bibitem[Midjourney.com(2022)]%
        {midjourney}
\bibfield{author}{\bibinfo{person}{Midjourney.com}.}
  \bibinfo{year}{2022}\natexlab{}.
\newblock \bibinfo{title}{Midjourney}.
\newblock
\newblock
\urldef\tempurl%
\url{https://midjourney.com/}
\showURL{%
\tempurl}
\newblock
\shownote{Retrieved on 3/15/2023.}.


\bibitem[Nichol et~al\mbox{.}(2022)]%
        {nichol2021glide}
\bibfield{author}{\bibinfo{person}{Alexander~Quinn Nichol},
  \bibinfo{person}{Prafulla Dhariwal}, \bibinfo{person}{Aditya Ramesh},
  \bibinfo{person}{Pranav Shyam}, \bibinfo{person}{Pamela Mishkin},
  \bibinfo{person}{Bob Mcgrew}, \bibinfo{person}{Ilya Sutskever}, {and}
  \bibinfo{person}{Mark Chen}.} \bibinfo{year}{2022}\natexlab{}.
\newblock \showarticletitle{GLIDE: Towards Photorealistic Image Generation and
  Editing with Text-Guided Diffusion Models}. In
  \bibinfo{booktitle}{\emph{International Conference on Machine Learning}}.
  PMLR, \bibinfo{pages}{16784--16804}.
\newblock


\bibitem[Ocvirk et~al\mbox{.}(1968)]%
        {ocvirk1968art}
\bibfield{author}{\bibinfo{person}{Otto~G Ocvirk}, \bibinfo{person}{Robert~E
  Stinson}, \bibinfo{person}{Philip~R Wigg}, \bibinfo{person}{Robert~O Bone},
  {and} \bibinfo{person}{David~L Cayton}.} \bibinfo{year}{1968}\natexlab{}.
\newblock \bibinfo{booktitle}{\emph{Art fundamentals: Theory and practice}}.
\newblock \bibinfo{publisher}{WC Brown Company}.
\newblock


\bibitem[OpenAI(2022)]%
        {chatgpt}
\bibfield{author}{\bibinfo{person}{OpenAI}.} \bibinfo{year}{2022}\natexlab{}.
\newblock \bibinfo{title}{ChatGPT}.
\newblock
\newblock
\urldef\tempurl%
\url{https://openai.com/blog/chatgpt}
\showURL{%
\tempurl}
\newblock
\shownote{Retrieved on 3/15/2023.}.


\bibitem[OpenAI(2023)]%
        {gpt4}
\bibfield{author}{\bibinfo{person}{OpenAI}.} \bibinfo{year}{2023}\natexlab{}.
\newblock \bibinfo{title}{GPT-4 Technical Report}.
\newblock
\newblock
\urldef\tempurl%
\url{https://cdn.openai.com/papers/gpt-4.pdf}
\showURL{%
\tempurl}
\newblock
\shownote{Retrieved on 3/15/2023.}.


\bibitem[Oppenlaender(2022a)]%
        {oppenlaender2022creativity}
\bibfield{author}{\bibinfo{person}{Jonas Oppenlaender}.}
  \bibinfo{year}{2022}\natexlab{a}.
\newblock \showarticletitle{The Creativity of Text-to-Image Generation}. In
  \bibinfo{booktitle}{\emph{Proceedings of the 25th International Academic
  Mindtrek Conference}}. \bibinfo{pages}{192--202}.
\newblock


\bibitem[Oppenlaender(2022b)]%
        {oppenlaender2022taxonomy}
\bibfield{author}{\bibinfo{person}{Jonas Oppenlaender}.}
  \bibinfo{year}{2022}\natexlab{b}.
\newblock \showarticletitle{A Taxonomy of Prompt Modifiers for Text-to-Image
  Generation}.
\newblock \bibinfo{journal}{\emph{arXiv preprint arXiv:2204.13988}}
  (\bibinfo{year}{2022}).
\newblock


\bibitem[Pavlichenko and Ustalov(2022)]%
        {pavlichenko2022best}
\bibfield{author}{\bibinfo{person}{Nikita Pavlichenko} {and}
  \bibinfo{person}{Dmitry Ustalov}.} \bibinfo{year}{2022}\natexlab{}.
\newblock \showarticletitle{Best Prompts for Text-to-Image Models and How to
  Find Them}.
\newblock  (\bibinfo{year}{2022}).
\newblock


\bibitem[Pressman et~al\mbox{.}(2022)]%
        {pressmancrowson2022}
\bibfield{author}{\bibinfo{person}{John~David Pressman},
  \bibinfo{person}{Katherine Crowson}, {and}
  \bibinfo{person}{Simulacra~Captions Contributors}.}
  \bibinfo{year}{2022}\natexlab{}.
\newblock \bibinfo{title}{Simulacra Aesthetic Captions}.
\newblock
\newblock
\urldef\tempurl%
\url{https://github.com/JD-P/simulacra-aesthetic-captions}
\showURL{%
\tempurl}
\newblock
\shownote{Retrieved on 3/15/2023.}.


\bibitem[Ramesh et~al\mbox{.}(2022)]%
        {ramesh2022hierarchical}
\bibfield{author}{\bibinfo{person}{Aditya Ramesh}, \bibinfo{person}{Prafulla
  Dhariwal}, \bibinfo{person}{Alex Nichol}, \bibinfo{person}{Casey Chu}, {and}
  \bibinfo{person}{Mark Chen}.} \bibinfo{year}{2022}\natexlab{}.
\newblock \showarticletitle{Hierarchical text-conditional image generation with
  clip latents}.
\newblock \bibinfo{journal}{\emph{arXiv preprint arXiv:2204.06125}}
  (\bibinfo{year}{2022}).
\newblock


\bibitem[Ramesh et~al\mbox{.}(2021)]%
        {ramesh2021zero}
\bibfield{author}{\bibinfo{person}{Aditya Ramesh}, \bibinfo{person}{Mikhail
  Pavlov}, \bibinfo{person}{Gabriel Goh}, \bibinfo{person}{Scott Gray},
  \bibinfo{person}{Chelsea Voss}, \bibinfo{person}{Alec Radford},
  \bibinfo{person}{Mark Chen}, {and} \bibinfo{person}{Ilya Sutskever}.}
  \bibinfo{year}{2021}\natexlab{}.
\newblock \showarticletitle{Zero-shot text-to-image generation}. In
  \bibinfo{booktitle}{\emph{International Conference on Machine Learning}}.
  PMLR, \bibinfo{pages}{8821--8831}.
\newblock


\bibitem[Reed et~al\mbox{.}(2016)]%
        {reed2016generative}
\bibfield{author}{\bibinfo{person}{Scott Reed}, \bibinfo{person}{Zeynep Akata},
  \bibinfo{person}{Xinchen Yan}, \bibinfo{person}{Lajanugen Logeswaran},
  \bibinfo{person}{Bernt Schiele}, {and} \bibinfo{person}{Honglak Lee}.}
  \bibinfo{year}{2016}\natexlab{}.
\newblock \showarticletitle{Generative adversarial text to image synthesis}. In
  \bibinfo{booktitle}{\emph{International conference on machine learning}}.
  PMLR, \bibinfo{pages}{1060--1069}.
\newblock


\bibitem[Rombach et~al\mbox{.}(2022)]%
        {rombach2022high}
\bibfield{author}{\bibinfo{person}{Robin Rombach}, \bibinfo{person}{Andreas
  Blattmann}, \bibinfo{person}{Dominik Lorenz}, \bibinfo{person}{Patrick
  Esser}, {and} \bibinfo{person}{Bj{\"o}rn Ommer}.}
  \bibinfo{year}{2022}\natexlab{}.
\newblock \showarticletitle{High-resolution image synthesis with latent
  diffusion models}. In \bibinfo{booktitle}{\emph{Proceedings of the IEEE/CVF
  Conference on Computer Vision and Pattern Recognition}}.
  \bibinfo{pages}{10684--10695}.
\newblock


\bibitem[Saharia et~al\mbox{.}(2022)]%
        {saharia2022photorealistic}
\bibfield{author}{\bibinfo{person}{Chitwan Saharia}, \bibinfo{person}{William
  Chan}, \bibinfo{person}{Saurabh Saxena}, \bibinfo{person}{Lala Li},
  \bibinfo{person}{Jay Whang}, \bibinfo{person}{Emily Denton},
  \bibinfo{person}{Seyed Kamyar~Seyed Ghasemipour}, \bibinfo{person}{Raphael
  Gontijo-Lopes}, \bibinfo{person}{Burcu~Karagol Ayan}, \bibinfo{person}{Tim
  Salimans}, {et~al\mbox{.}}} \bibinfo{year}{2022}\natexlab{}.
\newblock \showarticletitle{Photorealistic Text-to-Image Diffusion Models with
  Deep Language Understanding}. In \bibinfo{booktitle}{\emph{Advances in Neural
  Information Processing Systems}}.
\newblock


\bibitem[Sauer et~al\mbox{.}(2023)]%
        {sauer2023stylegan}
\bibfield{author}{\bibinfo{person}{Axel Sauer}, \bibinfo{person}{Tero Karras},
  \bibinfo{person}{Samuli Laine}, \bibinfo{person}{Andreas Geiger}, {and}
  \bibinfo{person}{Timo Aila}.} \bibinfo{year}{2023}\natexlab{}.
\newblock \showarticletitle{StyleGAN-T: Unlocking the Power of GANs for Fast
  Large-Scale Text-to-Image Synthesis}.
\newblock \bibinfo{journal}{\emph{arXiv preprint arXiv:2301.09515}}
  (\bibinfo{year}{2023}).
\newblock


\bibitem[Schuhmann et~al\mbox{.}(2021)]%
        {schuhmann2021laion}
\bibfield{author}{\bibinfo{person}{Christoph Schuhmann},
  \bibinfo{person}{Richard Vencu}, \bibinfo{person}{Romain Beaumont},
  \bibinfo{person}{Robert Kaczmarczyk}, \bibinfo{person}{Clayton Mullis},
  \bibinfo{person}{Aarush Katta}, \bibinfo{person}{Theo Coombes},
  \bibinfo{person}{Jenia Jitsev}, {and} \bibinfo{person}{Aran Komatsuzaki}.}
  \bibinfo{year}{2021}\natexlab{}.
\newblock \showarticletitle{Laion-400m: Open dataset of clip-filtered 400
  million image-text pairs}.
\newblock \bibinfo{journal}{\emph{arXiv preprint arXiv:2111.02114}}
  (\bibinfo{year}{2021}).
\newblock


\bibitem[Silverstein et~al\mbox{.}(1999)]%
        {silverstein1999analysis}
\bibfield{author}{\bibinfo{person}{Craig Silverstein}, \bibinfo{person}{Hannes
  Marais}, \bibinfo{person}{Monika Henzinger}, {and} \bibinfo{person}{Michael
  Moricz}.} \bibinfo{year}{1999}\natexlab{}.
\newblock \showarticletitle{Analysis of a very large web search engine query
  log}. In \bibinfo{booktitle}{\emph{Acm sigir forum}},
  Vol.~\bibinfo{volume}{33}. ACM New York, NY, USA, \bibinfo{pages}{6--12}.
\newblock


\bibitem[Sohl-Dickstein et~al\mbox{.}(2015)]%
        {sohl2015deep}
\bibfield{author}{\bibinfo{person}{Jascha Sohl-Dickstein},
  \bibinfo{person}{Eric Weiss}, \bibinfo{person}{Niru Maheswaranathan}, {and}
  \bibinfo{person}{Surya Ganguli}.} \bibinfo{year}{2015}\natexlab{}.
\newblock \showarticletitle{Deep unsupervised learning using nonequilibrium
  thermodynamics}. In \bibinfo{booktitle}{\emph{International Conference on
  Machine Learning}}. PMLR, \bibinfo{pages}{2256--2265}.
\newblock


\bibitem[Touvron et~al\mbox{.}(2023)]%
        {touvron2023llama}
\bibfield{author}{\bibinfo{person}{Hugo Touvron}, \bibinfo{person}{Thibaut
  Lavril}, \bibinfo{person}{Gautier Izacard}, \bibinfo{person}{Xavier
  Martinet}, \bibinfo{person}{Marie-Anne Lachaux},
  \bibinfo{person}{Timoth{\'e}e Lacroix}, \bibinfo{person}{Baptiste
  Rozi{\`e}re}, \bibinfo{person}{Naman Goyal}, \bibinfo{person}{Eric Hambro},
  \bibinfo{person}{Faisal Azhar}, {et~al\mbox{.}}}
  \bibinfo{year}{2023}\natexlab{}.
\newblock \showarticletitle{Llama: Open and efficient foundation language
  models}.
\newblock \bibinfo{journal}{\emph{arXiv preprint arXiv:2302.13971}}
  (\bibinfo{year}{2023}).
\newblock


\bibitem[Turc and Nemade(2022)]%
        {iuliaturc_gauravnemade_2022}
\bibfield{author}{\bibinfo{person}{Iulia Turc} {and} \bibinfo{person}{Gaurav
  Nemade}.} \bibinfo{year}{2022}\natexlab{}.
\newblock \bibinfo{title}{Midjourney User Prompts \& Generated Images (250k)}.
\newblock
\newblock
\urldef\tempurl%
\url{https://doi.org/10.34740/KAGGLE/DS/2349267}
\showDOI{\tempurl}
\newblock
\shownote{Retrieved on 3/15/2023.}.


\bibitem[Wang et~al\mbox{.}(2022)]%
        {wangDiffusionDBLargescalePrompt2022}
\bibfield{author}{\bibinfo{person}{Zijie~J Wang}, \bibinfo{person}{Evan
  Montoya}, \bibinfo{person}{David Munechika}, \bibinfo{person}{Haoyang Yang},
  \bibinfo{person}{Benjamin Hoover}, {and} \bibinfo{person}{Duen~Horng Chau}.}
  \bibinfo{year}{2022}\natexlab{}.
\newblock \showarticletitle{DiffusionDB: A Large-scale Prompt Gallery Dataset
  for Text-to-Image Generative Models}.
\newblock \bibinfo{journal}{\emph{arXiv preprint arXiv:2210.14896}}
  (\bibinfo{year}{2022}).
\newblock


\bibitem[Xie and O'Hallaron(2002)]%
        {xie2002locality}
\bibfield{author}{\bibinfo{person}{Yinglian Xie} {and} \bibinfo{person}{David
  O'Hallaron}.} \bibinfo{year}{2002}\natexlab{}.
\newblock \showarticletitle{Locality in search engine queries and its
  implications for caching}. In \bibinfo{booktitle}{\emph{Proceedings.
  Twenty-First Annual Joint Conference of the IEEE Computer and Communications
  Societies}}, Vol.~\bibinfo{volume}{3}. IEEE, \bibinfo{pages}{1238--1247}.
\newblock


\bibitem[Xie et~al\mbox{.}(2022)]%
        {xie2022much}
\bibfield{author}{\bibinfo{person}{Yutong Xie}, \bibinfo{person}{Ziqiao Xu},
  \bibinfo{person}{Jiaqi Ma}, {and} \bibinfo{person}{Qiaozhu Mei}.}
  \bibinfo{year}{2022}\natexlab{}.
\newblock \showarticletitle{How Much of the Chemical Space Has Been Explored?
  Selecting the Right Exploration Measure for Drug Discovery}. In
  \bibinfo{booktitle}{\emph{ICML 2022 2nd AI for Science Workshop}}.
\newblock


\bibitem[Xie et~al\mbox{.}(2023)]%
        {xie2023how}
\bibfield{author}{\bibinfo{person}{Yutong Xie}, \bibinfo{person}{Ziqiao Xu},
  \bibinfo{person}{Jiaqi Ma}, {and} \bibinfo{person}{Qiaozhu Mei}.}
  \bibinfo{year}{2023}\natexlab{}.
\newblock \showarticletitle{How Much Space Has Been Explored? Measuring the
  Chemical Space Covered by Databases and Machine-Generated Molecules}. In
  \bibinfo{booktitle}{\emph{The Eleventh International Conference on Learning
  Representations}}.
\newblock


\bibitem[Yang et~al\mbox{.}(2011)]%
        {yang2011query}
\bibfield{author}{\bibinfo{person}{Lei Yang}, \bibinfo{person}{Qiaozhu Mei},
  \bibinfo{person}{Kai Zheng}, {and} \bibinfo{person}{David~A Hanauer}.}
  \bibinfo{year}{2011}\natexlab{}.
\newblock \showarticletitle{Query log analysis of an electronic health record
  search engine}. In \bibinfo{booktitle}{\emph{AMIA annual symposium
  proceedings}}, Vol.~\bibinfo{volume}{2011}. American Medical Informatics
  Association, \bibinfo{pages}{915}.
\newblock


\bibitem[Yu et~al\mbox{.}(2022)]%
        {yu2022scaling}
\bibfield{author}{\bibinfo{person}{Jiahui Yu}, \bibinfo{person}{Yuanzhong Xu},
  \bibinfo{person}{Jing~Yu Koh}, \bibinfo{person}{Thang Luong},
  \bibinfo{person}{Gunjan Baid}, \bibinfo{person}{Zirui Wang},
  \bibinfo{person}{Vijay Vasudevan}, \bibinfo{person}{Alexander Ku},
  \bibinfo{person}{Yinfei Yang}, \bibinfo{person}{Burcu~Karagol Ayan},
  {et~al\mbox{.}}} \bibinfo{year}{2022}\natexlab{}.
\newblock \showarticletitle{Scaling Autoregressive Models for Content-Rich
  Text-to-Image Generation}.
\newblock \bibinfo{journal}{\emph{Transactions on Machine Learning Research}}
  (\bibinfo{year}{2022}).
\newblock


\bibitem[Zheng et~al\mbox{.}(2011)]%
        {zheng2011collaborative}
\bibfield{author}{\bibinfo{person}{Kai Zheng}, \bibinfo{person}{Qiaozhu Mei},
  {and} \bibinfo{person}{David~A Hanauer}.} \bibinfo{year}{2011}\natexlab{}.
\newblock \showarticletitle{Collaborative search in electronic health records}.
\newblock \bibinfo{journal}{\emph{Journal of the American Medical Informatics
  Association}} \bibinfo{volume}{18}, \bibinfo{number}{3}
  (\bibinfo{year}{2011}), \bibinfo{pages}{282--291}.
\newblock


\end{thebibliography}

\newpage
\appendix

\section{Data and Data Processing}
\label{app:data}

\subsection{Datasets}

For the datasets, we list important features (prompt, timestamp, user ID, and rating), feature descriptions, and corresponding examples in Table \ref{tab:dataset}.

\begin{table}[ht]
    \centering
    \caption{Feature descriptions and examples of the Midjourney, DiffusionDB, and SAC datasets.}
    \label{tab:dataset}
    \resizebox{\columnwidth}{!}{
    \begin{tabular}{c|>{\raggedright\arraybackslash}p{0.3\columnwidth}|>{\raggedright\arraybackslash}p{0.6\columnwidth}}
    \toprule
      & \textbf{Description} & \textbf{Examples} \\
    \midrule
   \multirow{3}{*}[-25pt]{\rotatebox[origin=c]{90}{\textbf{Prompt}}}  &  \multirow{3}{*}{\parbox{0.3\columnwidth}{\raggedright The prompt used to generate images. \newline \textbf{Type}: \texttt{String}}}   & \textbf{Midjourney}: \texttt{hands, by Karel Thole and Mike Mignola -{}-ar 2:3} \\
     &    &  \textbf{DiffusionDB}: \texttt{Ibai Berto Romero as Willy Wonka, highly detailed, oil on canvas} \\
     &    &  \textbf{SAC}: \texttt{concept art by David Production}.	  \\
    \midrule
   \multirow{3}{*}[-10pt]{\rotatebox[origin=c]{90}{\textbf{Timestamp}}}   &  \multirow{3}{*}{\parbox{0.3\columnwidth}{\raggedright The timestamp when the image was generated from the prompt. \newline \textbf{Type}: \texttt{String}}}   &  \textbf{Midjourney}: \texttt{2022-06-23T23:58:16.024000 +00:00} \\
     &    &  \textbf{DiffusionDB}: \texttt{2022-08-07 22: 57:00+00:00} \\
     &    &  \textbf{SAC}: N/A	  \\
    \midrule
   \multirow{3}{*}[-15pt]{\rotatebox[origin=c]{90}{\textbf{User ID}}}   &  \multirow{3}{*}{\parbox{0.3\columnwidth}{\raggedright The unique ID for the user account who submitted the prompt. \newline \textbf{Type}: \texttt{String}}}   & \textbf{Midjourney}: \texttt{977252506335858758} \\
     &    &  \textbf{DiffusionDB}: \texttt{fcdb3e09f977412c342b6624a19 d1295ee1334c153c90af16d1cca 8d9f27b04a} \\
     &    &  \textbf{SAC}: N/A	  \\
    \midrule
   \multirow{3}{*}[-5pt]{\rotatebox[origin=c]{90}{\textbf{Ratings}}}   &  \multirow{3}{*}{\parbox{0.3\columnwidth}{The rating of the image generated from the prompt\tablefootnote{Note that one prompt may correspond to multiple images, and one image may have multiple ratings. Here we list all the ratings correlated to the example prompt.}. \newline \textbf{Type}: \texttt{Integer}}}   & \textbf{Midjourney}: N/A \\
     &    &  \textbf{DiffusionDB}: N/A \\
     &    &  \textbf{SAC}: 6, 5, 7, 5 \newline	  \\
    \bottomrule
    \end{tabular}}
\end{table}


\subsection{Data Processing}

\paragraph{Midjourney}

We extracted prompts, timestamps, and user IDs from the records in the Midjourney dataset. The prompts in Midjourney may contain specific syntactic parameters of the Midjourney model, such as ``\texttt{-{}-ar}'' for aspect ratios, ``\texttt{-{}-h}'' for heights, ``\texttt{-{}-w}'' for widths, ``\texttt{::}'' for assigning weights to certain terms in the prompts.  
We first take the lowercase characters from tokenized prompts with the Spacy tokenizer\footnote{Spacy: \url{https://spacy.io/}, retrieved on 3/15/2023.}. 
Regarding the parameters, such as ``\texttt{-{}-h}'',  we consider them single terms. Specially, we split the weighted terms with their weights, and consider ``\texttt{::}'' and ``\texttt{::-}'' (negative weight) as two different terms.
During tokenization, we also removed redundant whitespaces. 
Midjourney allows users to upload reference images as parts of their prompts in the form of Discord links. These links are also processed as special terms. 

\paragraph{DiffusionDB} 

We utilize the metadata of DiffusionDB-Large (14M) for prompt analysis. We first remove duplicate data entries with the same prompt, timestamp, and user ID, meaning these entries record different images generated by the same user with the same prompt as a single submission. 
As a result, we obtained 2,208,019 non-duplication prompt submissions from users. Note that repeated submissions of prompts are reserved. 
We tokenize the prompts and remove the whitespace as we process the Midjourney data. 

\paragraph{SAC} SAC provides aesthetic ratings of generated images. Note that one prompt can correspond to multiple images, and each image can also have multiple ratings. Since there are no user ID and timestamp annotations in SAC, to remove the duplicates, we simply extract the unique prompts and conclude all correlated ratings. 


\bigbreak

More details can be found in the supplementary materials. 



\section{Additional Analysis Results}

\subsection{Prompt-Level Analysis}

\paragraph{\textbf{Prompt length distributions}}

\begin{figure}[ht]
  \centering
  \includegraphics[width=\linewidth]{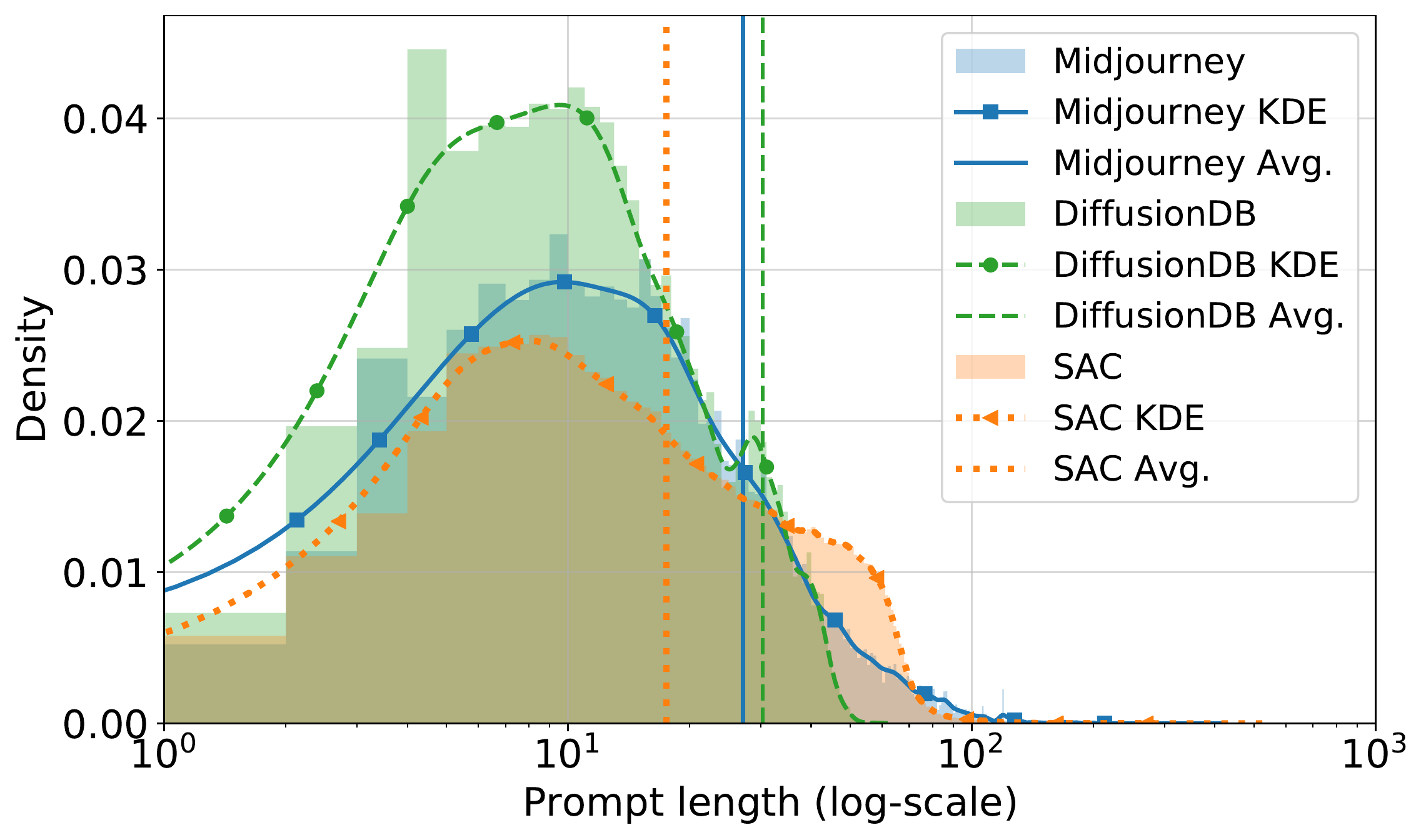}
  \vspace{-5pt}
  \caption{Prompt length distributions. The x-axis (prompt length) is plotted in the log scale. }
  \label{fig:prompt-lengths-hist}
\end{figure}

The distributions of prompt lengths are displayed in Figure \ref{fig:prompt-lengths-hist}, where the modes are around 10. 

\paragraph{\textbf{Prompts revised by users}}

Table \ref{tab:prompt-revisit} lists the most revisited prompts in DiffusionDB. 

\begin{table}[ht]
\caption{Most revisited prompts in DiffusionDB. Only revisits across sessions are considered. }
\label{tab:prompt-revisit}
\resizebox{\columnwidth}{!}{
\begin{tabular}{r|lr}
\toprule
  & \textbf{Prompt}                         & \hspace{-1cm}\textbf{\#Revisits} \\
\midrule
1  & test                                                                      & 24                                      \\
2  & cat                                                                       & 19                                      \\
3  & fat chuck is mad                                                          & 15                                      \\
4  &                                                                           & 15                                      \\
5  & dog                                                                       & 15                                      \\
6  & symmetry!! egyptian prince of technology, solid cube of light, ...        & 13                                      \\
7  & full character of a samurai, character design, painting by gaston ...     & 13                                      \\
8  & studio portrait of lawful good colorful female holy mecha paladin ...     & 11                                      \\
9  & full portrait and/or landscape. contemporary art print. high taste. ...   & 11                                      \\
10 & woman wearing oculus and digital glitch head edward hopper and ...        & 11                                      \\
11 & dream                                                                     & 10                                      \\
12 & hyperrealistic portrait of a character in a scenic environment by ...     & 10                                      \\
13 & full portrait \&/or landscape painting for a wall. contemporary art ...   & 10                                      \\
14 & zombie girl kawaii, trippy landscape, pop surrealism                      & 10                                      \\
15 & creepy ventriloquist dummy in the style of roger ballen, 4k, bw, ...      & 9                                       \\
16 & cinematic bust portrait of psychedelic robot from left, head and ...      & 9                                       \\
17 & red ball                                                                  & 9                                       \\
18 & amazing landscape photo of mountains with lake in sunset by ...           & 9                                       \\
19 & female geisha girl, beautiful face, rule of thirds, intricate outfit, ... & 9                                       \\
20 & full portrait and/or landscape painting for a wall. contemporary ...      & 9                                      \\
\bottomrule
\end{tabular}}
\end{table}

\paragraph{\textbf{Time series analysis}}

\begin{figure}[ht]
  \centering
  \includegraphics[width=\linewidth]{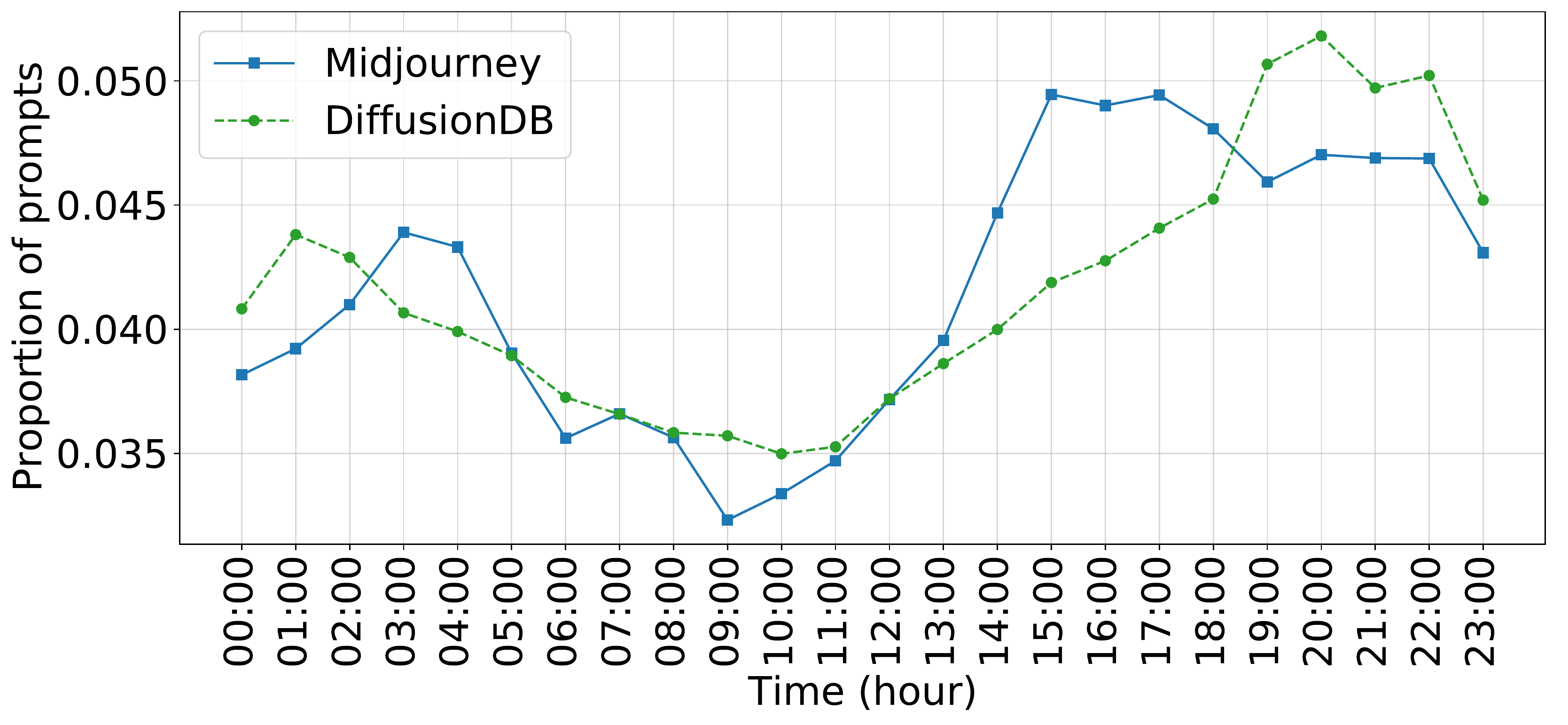}
  \vspace{-5pt}
  \caption{The distribution of prompts within 24 hours.}
  \label{fig:query-time-series}
  \vspace{-10pt}
\end{figure}

We analyze how the prompts distribute within 24 hours for the Midjourney and DiffusionDB datasets. The results are shown in Figure \ref{fig:query-time-series}. 
The patterns in these two datasets are similar: the rushing hours are around 01:00--03:00 (for both Midjourney and DiffusionDB), 15:00--17:00 (Midjourney), 20:00 (DiffusionDB); while during the daytime, the users are relatively inactive. 

\paragraph{\textbf{Ratings}} 

\begin{figure}[t]
  \centering
  \includegraphics[width=0.9\linewidth]{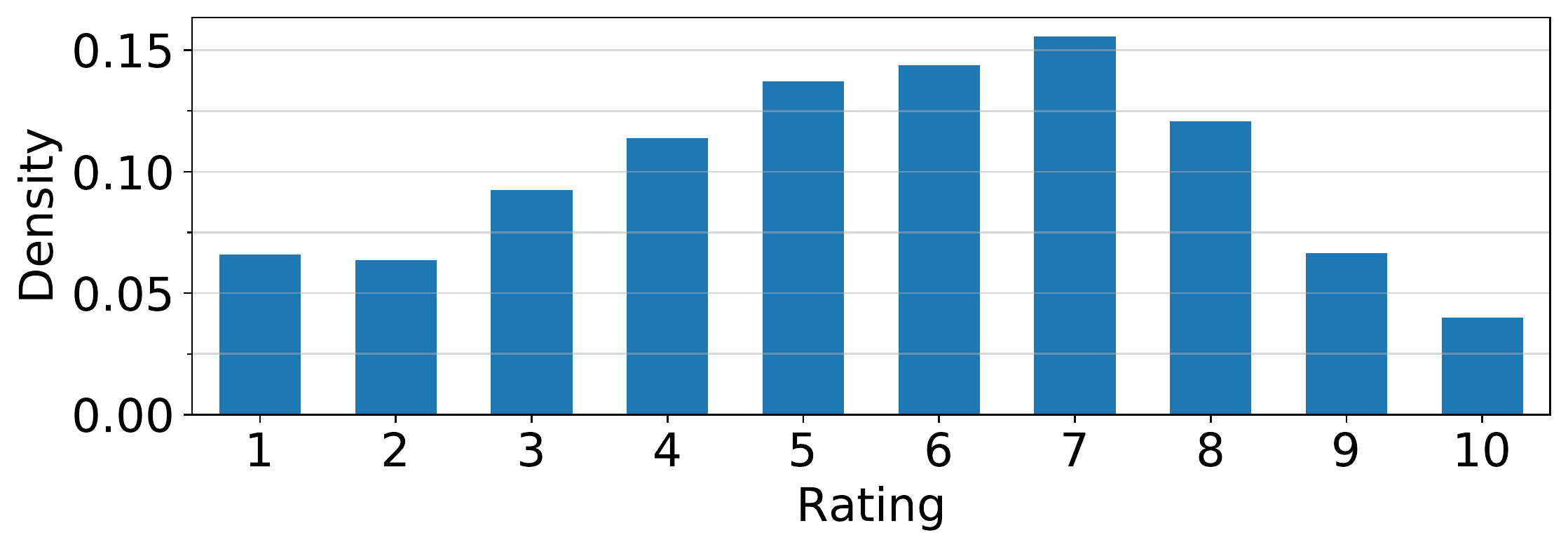}
  \vspace{-5pt}
  \caption{Rating distribution. Each rating corresponds to a user-input prompt and an image generated from that prompt. The average rating is 5.53, the standard deviation is 2.40, and the median is 6. }
  \label{fig:ratings-hist}
  \vspace{-10pt}
\end{figure}

The overall rating distribution of SAC is displayed in Figure \ref{fig:ratings-hist}. 

\begin{table}[ht]
\caption{Terms with the highest and the lowest average ratings. Only terms with frequencies larger than 100 are considered. ``Avg.'' and ``Std.'' are means and standard deviations of ratings respectively.}
\label{tab:terms-highest-lowest-ratings}
\vspace{-5pt}
\resizebox{\columnwidth}{!}{
\begin{tabular}{r|llll|llll}
\toprule
   & \multicolumn{4}{c|}{\textbf{Terms with highest avg. ratings}} & \multicolumn{4}{c}{\textbf{Terms with lowest avg. ratings}}       \\
\hhline{~--------}
   & \textbf{Term}             & \textbf{Avg.}      & \textbf{Std.}     & \textbf{Freq.}     & \textbf{Term}             & \textbf{Avg.}     & \textbf{Std.}     & \textbf{Freq.}     \\
\midrule
1  & shinjuku         & 8.55      & 0.90     & 168       & equations        & 2.36     & 2.18     & 240       \\
2  & gyuri            & 8.22      & 1.65     & 219       & mathematical     & 2.37     & 2.18     & 230       \\
3  & lohuller         & 8.22      & 1.66     & 215       & geismar          & 2.67     & 2.13     & 136       \\
4  & afremov          & 7.95      & 1.73     & 288       & haviv            & 2.68     & 2.14     & 136       \\
5  & leonid           & 7.95      & 1.73     & 288       & chermayeff       & 2.73     & 2.14     & 136       \\
6  & retrofuture      & 7.95      & 1.97     & 307       & learning         & 3.10     & 2.64     & 112       \\
7  & merantz          & 7.93      & 1.77     & 463       & pegasus          & 3.10     & 2.00     & 129       \\
8  & josan            & 7.91      & 1.73     & 1,647     & teacher          & 3.11     & 2.59     & 110       \\
9  & fantasyland      & 7.90      & 1.52     & 114       & someone          & 3.14     & 2.45     & 574       \\
10 & gensokyo         & 7.89      & 1.34     & 281       & funny            & 3.17     & 2.52     & 208       \\
\bottomrule
\end{tabular}}
\vspace{-10pt}
\end{table}

\subsection{Comparing Prompts with Training Data}
\label{app:laion}


To compare user-input prompts with texts that are used to train the text-to-image generation models, we also include the LAION dataset \cite{schuhmann2021laion}. 
LAION is a public dataset of CLIP-filtered image-text pairs and has often been used in large text-to-image model training \cite{rombach2022high,saharia2022photorealistic,yu2022scaling,sauer2023stylegan}. In the analysis, we use the LAION-400M dataset\footnote{LAION-400M dataset: \url{https://laion.ai/blog/laion-400-open-dataset/}. } that contains only English texts.




\paragraph{\textbf{Visualization}}

\begin{figure}[t]
  \centering
  \includegraphics[width=0.8\linewidth]{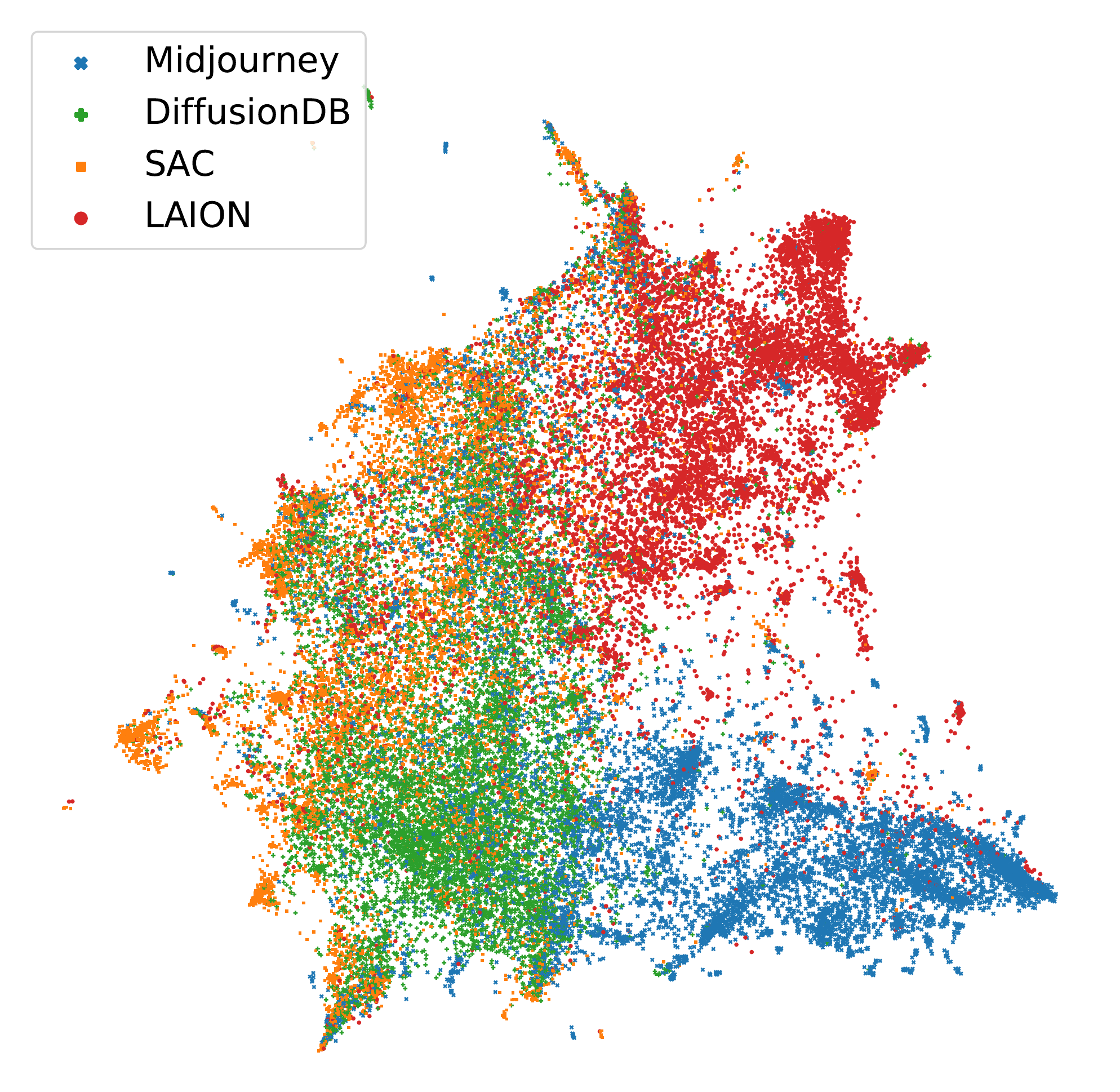}
  \vspace{-5pt}
  \caption{UMAP visualization of prompt embeddings. A clear gap can be identified between the LAION training data (red circles) and the user-input prompts (other colors). }
  \label{fig:prompt-emb}
\end{figure}

To intuitively see how user-input prompts and texts from the LAION training set are distributed, we use UMAP \cite{mcinnes2018umap} to visualize the prompts and the texts based on BERT \cite{devlin2018bert} embeddings in Figure \ref{fig:prompt-emb}. In the visualization, we find a clear gap between LAION (red circles) and other datasets, meaning the training set can hardly represent the real data distributions of user-input prompts. 
This visualization also aligns with the findings about vocabulary coverage, where we discover the terms in SAC are most covered by LAION, and the vocabulary of Midjorney is most distant from that of LAION. 

\paragraph{\textbf{Non-representative training data}}

We discover a huge gap between the user-input prompts and the texts in the open training data such as the LAION training set. 
The out-of-vocabulary (OOV) problem is severe, and in the prompts from Midjorney, about 75\% terms are not covered by LAION's vocabulary. Figure \ref{fig:prompt-emb} also displays a gap in prompt embedding distributions. 
All this evidence proves that the texts (mostly image captions) from the open training data can hardly represent users' information needs and we should call for another way that renders better supervision during training. 
ChatGPT \cite{chatgpt} has already demonstrated that reinforcement learning from human feedback (RLHF) \cite{christiano2017deep} could provide rich supervision and guidance to the model. 
However, for text-to-image generation, related work is still limited. Note that our analysis is based on the open datasets that are included in the training data of the models and doesn't consider the private training data that could have a different coverage of the space.

\end{document}